\pdfoutput=1
\documentclass[12pt,draftclsnofoot,journal,onecolumn]{IEEEtran}
\usepackage{amsmath,amsfonts}
\usepackage{algorithmic}
\usepackage{algorithm}
\usepackage{multirow}
\usepackage{multicol}
\usepackage{booktabs}
\usepackage{array}
\usepackage{units}
\usepackage{svg}
\usepackage[caption=false,font=small,labelfont=rm,textfont=rm,justification=centering,singlelinecheck=off]{subfig}
\usepackage{caption}
\usepackage{float}
\usepackage{textcomp}
\usepackage{stfloats}
\usepackage{url}
\usepackage{verbatim}
\usepackage{graphicx}
\usepackage{cite}
\begin{document}

\title{A Physical-Based Perspective for Understanding and Utilizing Spatial Resources of Wireless Channels}

\author{Hui Xu, Jun Wei Wu,~\IEEEmembership{Member,~IEEE}, Zhen Jie Qi,
Hao Tian Wu, Rui Wen Shao, Jieao Zhu, Qiang Cheng,~\IEEEmembership{Senior Member,~IEEE},
Linglong Dai,~\IEEEmembership{Fellow,~IEEE}, \\and
Tie Jun Cui,~\IEEEmembership{Fellow,~IEEE}
\thanks{This work was supported by the National Natural Science Foundation of China under Grants 62288101, 62171124, and 62225108; Jiangsu Province Frontier Leading Technology Basic Research Project under Grant BK20212002; the 111 Project under Grant 111-2-05; the Fundamental Research Funds for the Central Universities under Grants 2242023k5002 and 2242024k30009. \textit{(Hui Xu and Jun Wei Wu contributed equally to this work and are the co-first authors.) (Corresponding author: Tie Jun Cui)}}
\thanks{Hui Xu, Jun Wei Wu, Zhen Jie Qi, Rui Wen Shao, Qiang Cheng, and Tie Jun Cui are with the State Key Laboratory of Millimeter Waves, Southeast University, Nanjing 210096. Hui Xu is also with Zhangjiang Laboratory, Shanghai 201210, China; Jun Wei Wu is also with Pazhou Laboratory (Huangpu), Guangzhou 510555, China; and Tie Jun Cui is also with Suzhou Laboratory, Suzhou 215000, China (email: hui\_x@seu.edu.cn, jwwu@seu.edu.cn, 230239408@seu.edu.cn, srwsrw@126,com, qiangcheng@seu.edu.cn, and tjcui@seu.edu.cn).}
\thanks{Hao Tian Wu is with the School of Electrical and Electronic Engineering, Nanyang Technological University, 50 Nanyang Avenue, Singapore 639798, Singapore (email: haotian.wu@ntu.edu.sg).}
\thanks{Jieao Zhu and Linlong Dai is with Beijing National Research Center for Information Science and
Technology and Department of Electronic Engineering, Tsinghua University, Beijing 100084, China (e-mail: zja21@mails.tsinghua.edu.cn and daill@tsinghua.edu.cn).}}


\maketitle

\begin{abstract}To satisfy the increasing demands for transmission rates of wireless communications, it is necessary to use spatial resources of electromagnetic (EM) waves. In this context, EM information theory (EIT) has become a hot topic by integrating the theoretical framework of deterministic mathematics and stochastic statistics to explore the transmission mechanisms of continuous EM waves. However, the previous studies were primarily focused on frame analysis, with limited exploration of practical applications and a comprehensive understanding of its essential physical characteristics. In this paper, we present a three-dimensional (3-D) line-of-sight channel capacity formula that captures the vector EM physics and accommodates both near- and far-field scenes. Based on the rigorous mathematical equation and the physical mechanism of fast multipole expansion, a channel model is established, and the finite angular spectral bandwidth feature of scattered waves is revealed. To adapt to the feature of the channel, an optimization problem is formulated for determining the mode currents on the transmitter, aiming to obtain the optimal design of the precoder and combiner. We make comprehensive analyses to investigate the relationship among the spatial degree of freedom, noise, and transmitted power, thereby establishing a rigorous upper bound of channel capacity. A series of simulations are conducted to validate the theoretical model and numerical method. This work offers a novel perspective and methodology for understanding and leveraging EIT, and provides a theoretical foundation for the design and optimization of future wireless communications. 
\end{abstract}


\section{Introduction}
\IEEEPARstart{A}{s} a pioneer of information theory, Shannon discovered the transmission limit for noisy channels, and theoretically proved the existence of a certain communication strategy to achieve this limit \cite{ref1}. The maximum capacity of information transmission reads: $C=Blog_2(1+P/N)$, where $B$ is the signal bandwidth, $P$ is the transmitting power, and $N$ is the noise power \cite{ref1}. The Shannon theory has guided the development of both wired and wireless communications. Nowadays, as the maximal power is limited and the spectrum is overcrowded, the channel capacity of the fourth-generation communication systems has approached the Shannon limit \cite{ref2}. Great success of massive multi-input and multi-output (MIMO) in the fifth- generation communication makes spatial division multiplexing access (SDMA) a reality \cite{ref3,ref4}. Inspired by the massive MIMO, the spatial resources of electromagnetic (EM) waves have attracted ever-growing attention to meet the increasing demands for broad bandwidth, low latency, and high security in the next-generation networks \cite{ref5}. The emerging technologies include holographic MIMO \cite{ref6,ref7}, extremely large antenna arrays \cite{ref8}, near-field communications \cite{ref9,ref10}, programmable metasurfaces and reconfigurable intelligent surfaces (RIS) \cite{ref11,ref12,ref13}. To comprehensively explore the capabilities and limits of these technologies, it is essential to develop a unified framework that integrates the theories of deterministic physics and statistical mathematics to study the mechanisms of information transmissions in 
spatially continuous EM fields, leading to the concept of so-called EM information theory (EIT) \cite{ref14,ref15,ref16}.

Historically speaking, the integration of information and EM wave theory has been extensively explored since Shannon’s milestone paper. The earliest work of Gabor indicated the intrinsic connection between the transmission of information and physical states \cite{ref17}. With the introduction of Slepian power concentration functions over a finite interval \cite{ref18,ref19}, Francia undertook rigorous development of the information theory for image systems, laying the foundation to the well-known discipline of information optics \cite{ref20,ref21}. Bucci demonstrated that the EM fields radiated by scattering systems are essentially spatially bandlimited \cite{ref22}, and thus were represented on an arbitrary observation region by a linear combination of a finite number of basic orthogonal functions \cite{ref23}. The number of the bases is associated to the number of available parallel channels, leading to the concept of degrees of freedom (DoF). Then, the analysis of spatial DoF of different radiative systems has emerged as a central topic in developing the EIT \cite{ref24}.

There are three principal methodologies to study the spatial DoF to date, namely, the Hilbert-Schmidt decomposition of the compact Green’s function operator \cite{ref25,ref26,ref27,ref28,ref29}, the cut-set integral \cite{ref30,ref31,ref32} and the Kolmogorov information theory (KIT) \cite{ref33,ref34}. The method of the Hilbert-Schmidt decomposition is to solve dual eigenequations with the Green’s operator serving as the integral kernel, which enables the determination of available spatial channels and the coupling strengths. However, this method needs the complete channel state information and features high computational complexity. The cut-set integral is a low-complexity method to estimate the spatial DoF for arbitrary radiators, which characterizes the geometric variation of information flow over the observable region. Nevertheless, it cannot provide practical strategies for implementing the spatial communications. KIT can transform the information transmission into the mapping of a high- dimensional space, analogous to the intricate packing of hyperspheres. Although the physical insight of the KIT method is aesthetically pleasing, the mathematical problem of hypersphere packing remains a formidable challenge, yet to be resolved. 

Despite the extensive history EIT has amassed, the research interests continue expanding in the 2020s, with a notable influx of new technological developments occurring concurrently. The current representative work is not limited to the analysis of the spatial DoF. Rather, it encompasses several additional topics, including the fundamental capacity bounds of the communication systems \cite{ref34,ref35,ref36,ref37,ref38,ref39}, the physical limit of a radiator \cite{ref40,ref41}, novel consistent channel models \cite{ref42}, system architectures and applications \cite{ref43,ref44}, and other related subjects. Among these topics, the research on the continuous-space information capacity is crucial. Based on KIT, Migliore provided the physical interpretation of spatial DoF and derived a general EM capacity formula for spatial-temporal radiation systems \cite{ref34}. Jensen considered the information capacity of three-dimensional (3-D) continuous space under the constraint of finite radiation power \cite{ref35}, while Jeon extended the work to the transmitter and receiver with medium loss \cite{ref36}. Franceschetti proposed the spatial-temporal duality principle and obtained an EM capacity formula similar to the work of Migliore \cite{ref37}. The recent work of Dai. et al. investigated the capacity between continuous transmitters and receivers with finite size based on random field theory \cite{ref38}, and the analysis on MIMO capacity \cite{ref39}. Upon analysis of the above works, it was found that the existing EM capacity formulas predominantly take a similar form of $Nlog(1+P/N\sigma^2)$ , where $N$ represents the spatial DoF. The majority of these theories are predicated on finite power constraints and noise models between the transmitter and the receivers. Consequently, there has been comparatively little work that has comprehensively elucidated the information capacity formula from the perspective of the underlying physical characteristics, namely, the finite angular spectrum of scattered waves, which represents a knowledge gap that needs to be addressed in the theoretical development.

In this paper, we derive a 3-D EM channel capacity formula that quantifies the maximal number of distinguishable radiation fields between the transmitter and receiver of given apertures and distance. The main contributions of this work are summarized as follows.

Firstly, starting from rigorous mathematical equations of EM fields, we employ the fast multipole method (FMM) to construct a physical-consistent channel model that encompasses all-vector EM waves, including both scenes of near and far fields. By conducting an in-depth analysis of FMM, we reveal the finite angular spectral characteristics of the channel model. The model demonstrates that the interaction between the field and source can be effectively described by Fourier plane-wave theory, and is primarily embodied by wavenumber vectors within the finite angular spectral region.

Subsequently, we formulate an optimization problem of mode currents on the transmitter that support the maximal power of transmission between the transmitting and receiving apertures. By solving this problem, we obtain the optimal design of precoder and combiner in the spatial domain, thereby constructing a series of parallel independent subchannels. Further, we conduct a comprehensive analysis on the relationship between the spatial DoF, noise, and transmitted power, thereby establishing a rigorous bound of the channel capacity.

Finally, we conduct several numerical experiments to verify the effectiveness of our theory and the value of comprehending and leveraging EIT, focusing on the finite angular spectral characteristics of the channel, the intrinsic features of the communication system (i.e., eigenvalues and eigenfunctions), and the channel capacity.

\textit{Notation:} In this paper, bold calligraphy letters, such as $\boldsymbol{\mathcal{X}}$ and $\boldsymbol{\mathcal{Y}}$, denote the sets or Hilbert space, and $|\boldsymbol{\mathcal{X}}|$ represents the Lebesgue measure of $\boldsymbol{\mathcal{X}}$; boldface uppercase $\mathbf{A}$ and lowercase $\mathbf{a}$ letters denote the matrix and vector, respectively; $|\mathbf{a}|$ denotes the $l_2$ norm of $\mathbf{a}$, and $\mathbf{\hat{a}}=\mathbf{a}/|\mathbf{a}|$ represents its unit direction vector; the character $\mathbf{\bar{A}}$ denotes the dyadic; the complex character is $j=\sqrt{-1}$; the dot $\cdot$ denotes the scalar product of two vectors, or the matrix-vector multiplication; $\mathbb{E}[\mathbf{x}]$ denotes the mean of random variable $\mathbf{x}$; $\mathbf{a}^T$ denotes the transpose of $\mathbf{a}$, $\mathbf{\bar{a}}$  and $\bar{f}(x)$ represent their conjugates; $*$ denotes the convolution operation, $\mathcal{F}[f(x)]$ and $\mathcal{F}^{-1}[f(x)]$ are the Fourier and inverse transforms of $f(x)$; $\delta_{kl}$ denotes the Kronecker function, where $\delta_{kl}=1$ if $k=l$ and $\delta_{kl}=0$ otherwise. Finally, following the convention of engineering EM, time dependence $exp(j2{\pi}ft)$ ($f$ is the frequency of field) is exploited and dropped.
\begin{figure}[!t]
\centering
\subfloat[]{\includegraphics[width=0.9\textwidth,height=5cm]{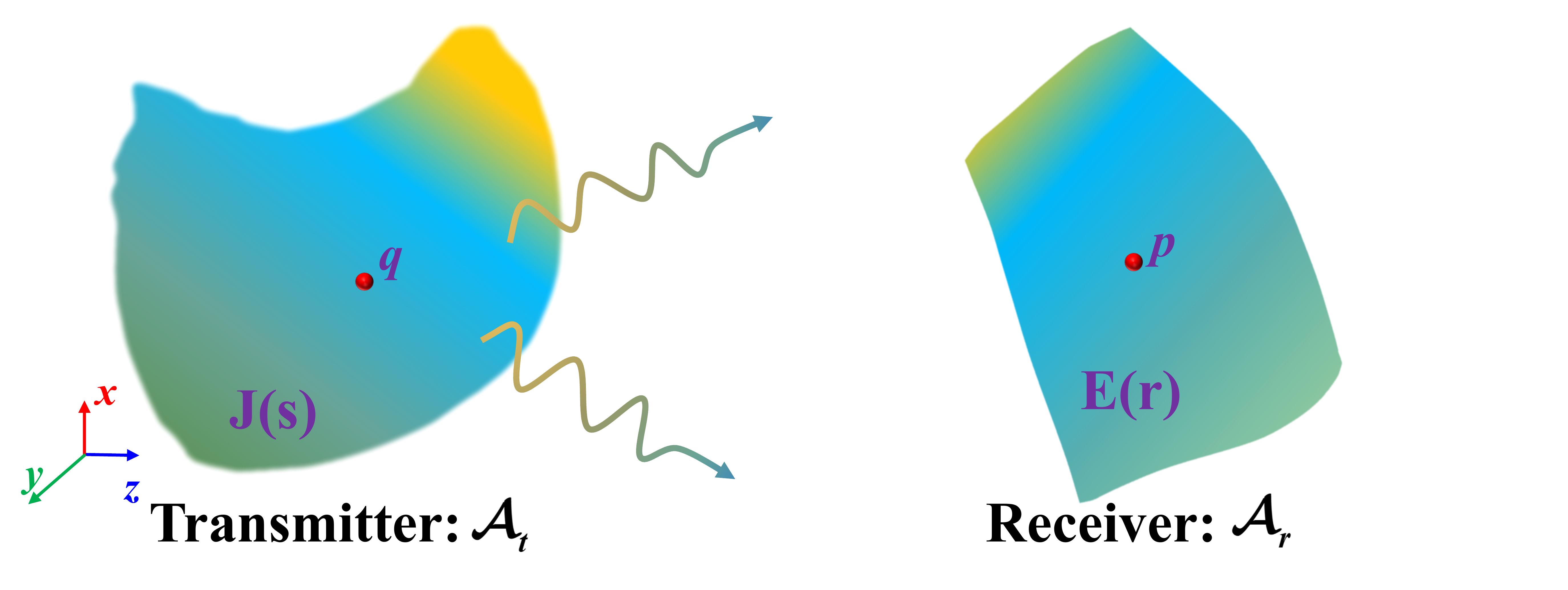}
\label{fig1a}}%
\hfil
\subfloat[]{\includegraphics[width=0.9\textwidth]{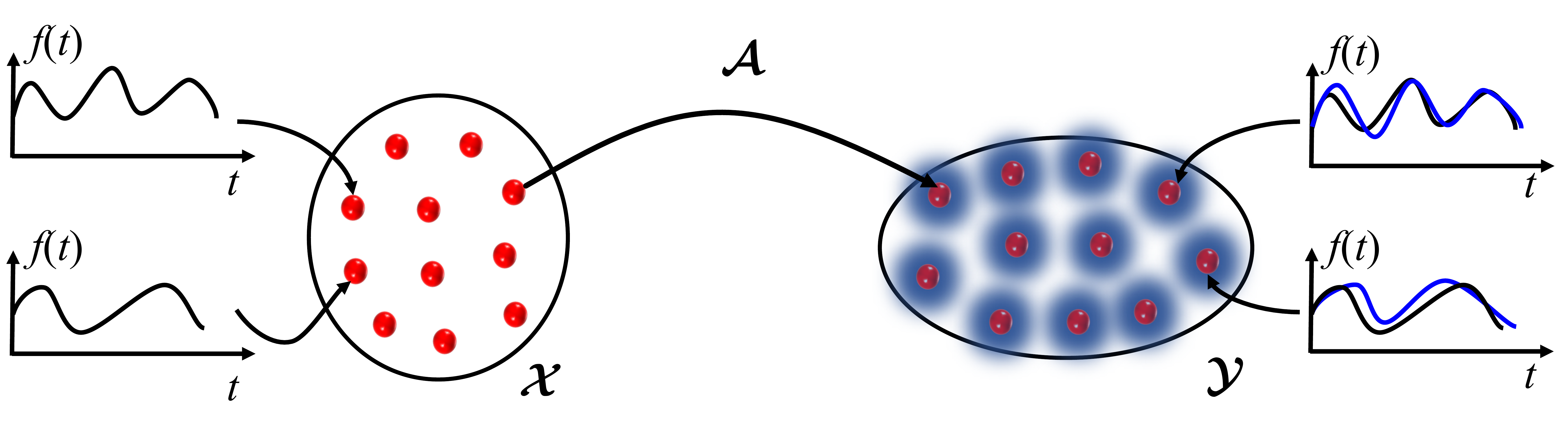}
\label{fig1b}}%
\hfil
\subfloat[]{\includegraphics[width=0.9\textwidth]{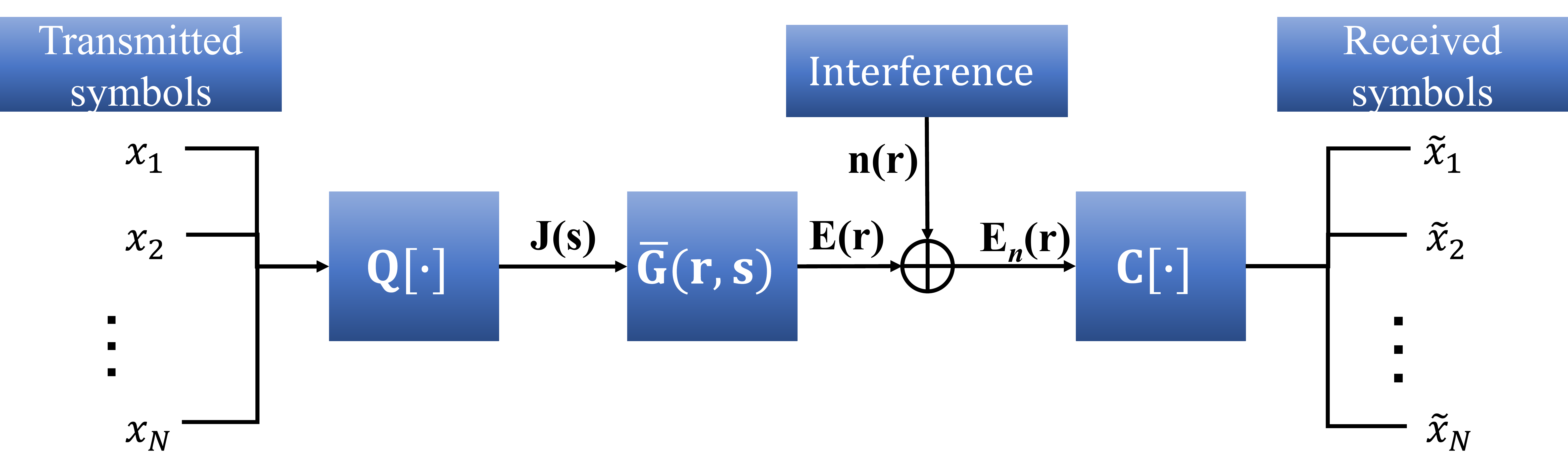}\label{fig1c}}%
\caption{Three EM channel analysis models for understanding the mechanism of wireless information transmission. (a) Physical representation of the EM channel. (b) Geometric interpretation of the transmission based on KIT. The black curves on the left represent the transmitted signals, while the blue curves on the right denote the received signals distorted by noises. (c) Block diagram of the system structure for illustrating the mechanism of the spatial information transmission.}
\label{fig1}
\end{figure}
\section{System Model and Problem Formulation}
Fig. \ref{fig1} presents the mechanism of information transmission using spatial resources. A representative free-space communication scenario is shown in Fig. \ref{fig1}\subref{fig1a}, where $\boldsymbol{\mathcal{A}_t}$ and $\boldsymbol{\mathcal{A}_r}$, which possess arbitrary 3-D shapes and finite apertures, serve as the transmitter and receiver, respectively. Fig. \ref{fig1}\subref{fig1b} presents the geometric interpretation of the wireless information transmission in the presence of noise based on KIT. Due to the physical constraint, the source current and the received field is confined to the compact Hilbert spaces $\boldsymbol{\mathcal{X}}$ and $\boldsymbol{\mathcal{Y}}$, respectively \cite{ref45}. From the perspective of the transmitter, each current distribution refers to a hypersphere in  and represents a unique signal. In the presence of noise, the corresponding field becomes a hypersphere with a shadow region of radius $\varepsilon$ in $\boldsymbol{\mathcal{Y}}$. Intuitively, reliable communications require that the shaded regions of any two balls do not overlap. In this context, the maximal amount of information that can be conveyed from the transmitter to the receiver is the number of hyperspheres with radius $\varepsilon$ that can be packed into the space $\boldsymbol{\mathcal{Y}}$. However, the problem is mathematically well-known as the optimal hypersphere packing in high-dimensional space, which is extremely hard and counterintuitive.

Fig. \ref{fig1}\subref{fig1c} illustrates the system structure of the spatial information transmission, which extends the classic vertical Bell Labs space-time (V-BLAST) MIMO architecture under the continuous-space EM channel \cite{ref46}. Based on the architecture, it is anticipated that the EM channel capacity formula will be deduced with the assistance of MIMO theory, thus overcoming the difficulty of KIT. In the process of information transmission, the first step involves the combination of multiple independent constellation symbols $\{x_n\}_{n=1}^N$ into a unique current through the precoder $\mathbf{Q[\cdot]}$. Subsequently, the receiver performs the combiner $\mathbf{C[\cdot]}$ on the signal containing noise to extract the multiple information hidden in the received field. It is worth noting that the precoder and combiner represent orthogonal bases in spaces $\boldsymbol{\mathcal{X}}$ and $\boldsymbol{\mathcal{Y}}$, respectively. To maximize the system capacity, these functions are designed according to the geometry of the transmitter and receiver. Recently, the precoder and combiner were obtained by the Fourier plane-wave approximation and optimization technology, thereby enabling the realization of wavelength division multiplexing (WDM) and pattern division multiplexing (PDM) of spatially continuous EM channels \cite{ref47,ref48}. However, the methodology presented here differs from previous work. By searching for the current distribution that maximizes the power transmission between the transmitter and receiver, we obtain a series of orthogonal bases, i.e., the explicit forms of the precoder and combiner.

\subsection{System Model}
\begin{figure*}[!t]
\centering
\subfloat[]{\includegraphics[width=0.9\textwidth]{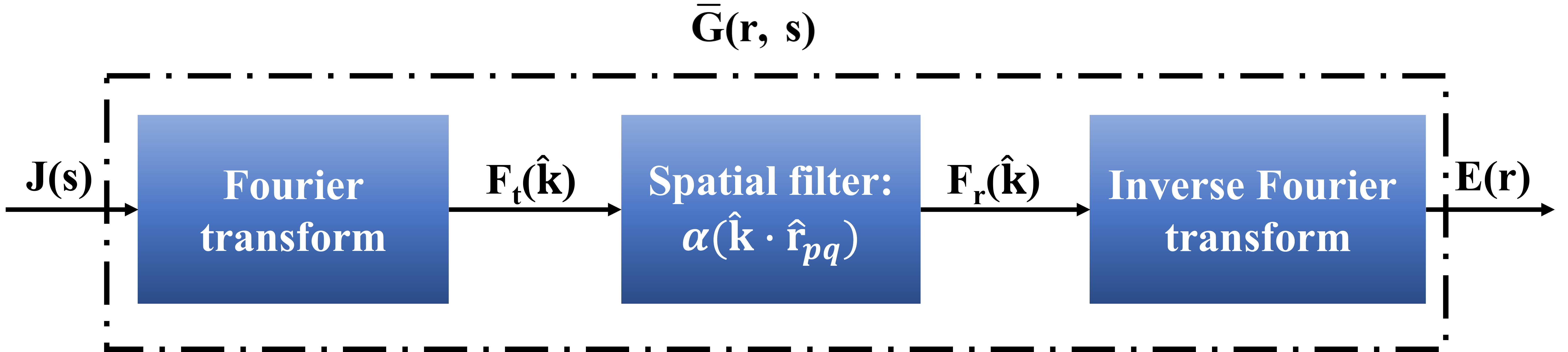}\label{fig2a}}%
\hfil
\subfloat[]{\includegraphics[width=0.45\textwidth]{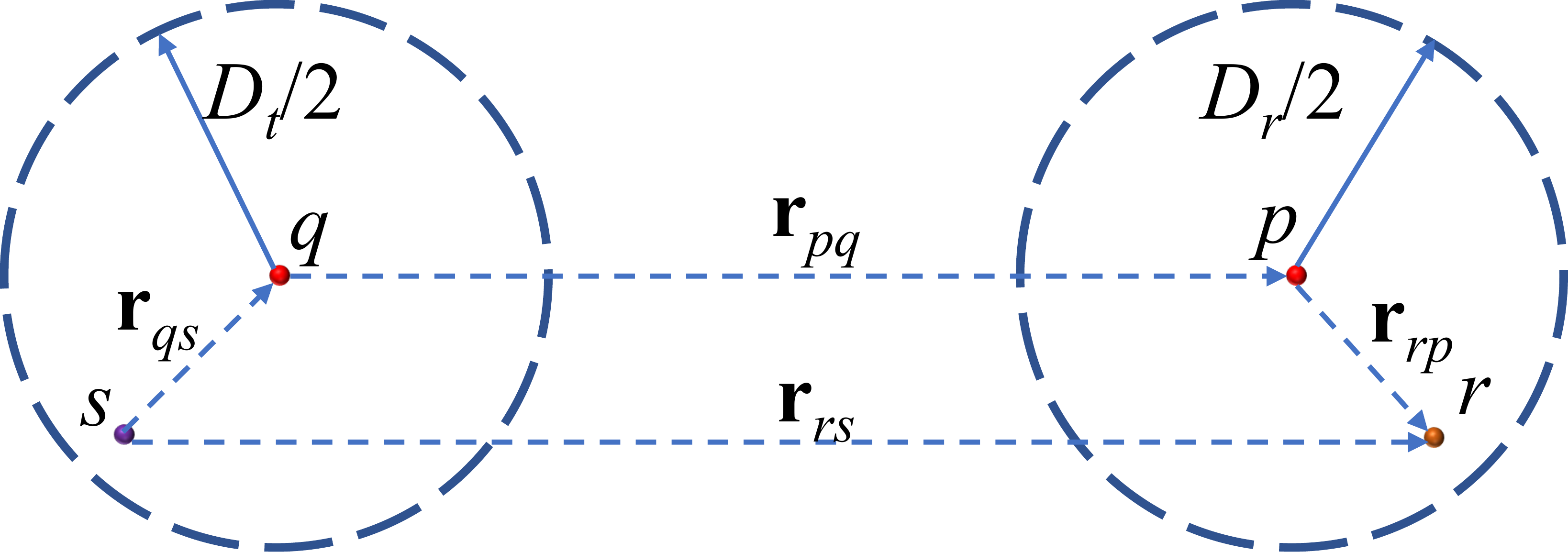}\label{fig2b}}%
\hfill
\subfloat[]{\includegraphics[width=0.45\textwidth]{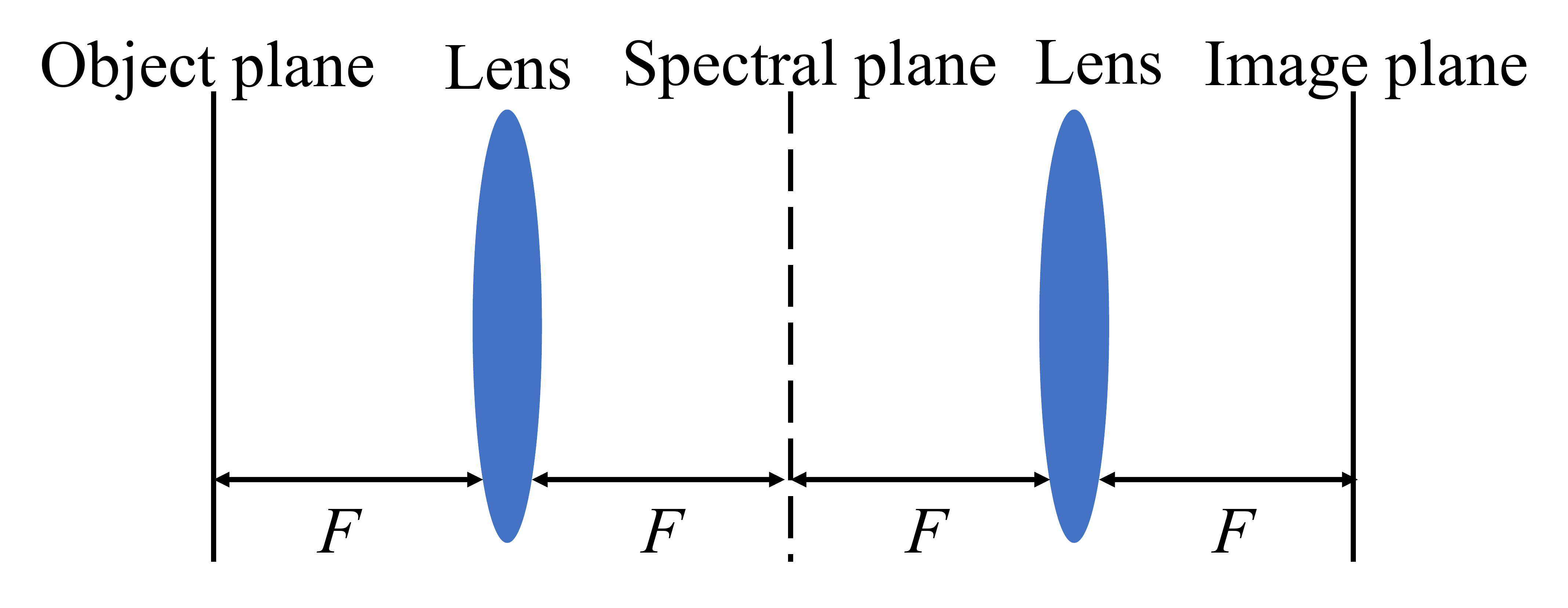}\label{fig2c}}%
\caption{Diagrams for understanding \eqref{eq4} and \eqref{eq6}, and the propagation operator. (a) Block diagram for illustrating the propagation operator. (b) Geometric coordinates for depicting the additional theorem expansion of SGF, namely, the equation \eqref{eq4}. (c) The classic 4-f system as a physical mapping of the propagation operator.}
\label{fig2}
\end{figure*}

The electric field   on the receiver radiated by the source is expressed as
\begin{equation}
\label{eq1}
\mathbf{E(r)} = -j\omega \mu\int_{\boldsymbol{\mathcal{A}_t}} \mathbf{\bar{G}(r,\;s)} \cdot \mathbf{J(s)}d\mathbf{s},\quad \mathbf{r}\in \boldsymbol{\mathcal{A}_t}.
\end{equation} The expression of dyadic Green’s function (DGF) in free space is \cite{ref49,ref50}
\begin{equation}
\label{eq2}
\begin{split}
    \mathbf{\bar{G}(r,\;s)} &=\left[\mathbf{\bar{I}}+\frac{\nabla\nabla}{k^2}\right]g(\mathbf{r,\,s}) \\
        &= \left[(\mathbf{\bar{I}}-\mathbf{\hat{R}\hat{R}})-\frac{j}{kR}(\mathbf{\bar{I}}-3\mathbf{\hat{R}\hat{R}})-\frac{1}{k^2R^2}(\mathbf{\bar{I}}-3\mathbf{\hat{R}\hat{R}})\right] g(\mathbf{r,\,s}),
\end{split}
\end{equation} where $\mathbf{\bar{I}}$ is the unit dydic, $\mathbf{r}$ and $\mathbf{s}$ are the positions of source and field points, respectively, $\mathbf{R=r-s}$, $R=\vert\mathbf{r-s}\vert$, and $g(\mathbf{r,\,s})=\nicefrac{e^{-jkR}}{4{\pi}R}$ is the free-space scalar Green’s function (SGF). In the explicit expression of DGF, the first term contributes significantly since the power of the other two terms decays to a negligible strength in a few wavelengths away from the source. Using the approximation $\mathbf{\hat{k}=\hat{r}\approx\hat{R}}$, the DGF is then degenerated to
\begin{equation}
\label{eq3}
     \mathbf{\bar{G}(r,\;s)} \approx (\mathbf{\bar{I}-\hat{k}\hat{k}})\frac{e^{-jk\vert\mathbf{r-s}\vert}}{4{\pi}\vert\mathbf{r-s}\vert}.
\end{equation}
When the distance between the centers of the transmitter and receiver is greater than the sum of the apertures\footnote{We should remark that the scenes within the distance are still near-field communications in most configurations of the transmitter and receiver.}, SGF can be expanded using the addition theorem \cite{ref50,ref51,ref52}
\begin{equation}
\label{eq4}
    \frac{e^{-jk\vert\mathbf{r-s}\vert}}{4{\pi}\vert\mathbf{r-s}\vert}=
    \frac{-jk}{16\pi^2}\oint_{S_E} e^{-jk\mathbf{\hat{k}}\cdot\mathbf{r}_{qs}}\cdot \alpha(\mathbf{\hat{k}}\cdot\mathbf{\hat{r}}_{pq})\cdot e^{-jk\mathbf{\hat{k}}\cdot\mathbf{r}_{rp}} d\mathbf{\hat{k}},
\end{equation} where the integral is performed on the unit sphere $S_E$. Fig. \ref{fig2}\subref{fig2b} shows the geometric coordinates for illustrating the above equation, where $\mathbf{r}_q$ and $\mathbf{r}_p$ are the centers of source and field regions, respectively; $\mathbf{r}_{qs}=\mathbf{r}_q-\mathbf{r}_s$, $\mathbf{r}_{pq}=\mathbf{r}_p-\mathbf{r}_q$, $\mathbf{r}_{rp}=\mathbf{r}_r-\mathbf{r}_q$; and $\alpha(\mathbf{\hat{k}}\cdot\mathbf{\hat{r}}_{pq})$ is the translator between the transmitter and receiver, namely,

\begin{equation}
\label{eq5}
\alpha(\mathbf{\hat{k}}\cdot\mathbf{\hat{r}}_{pq})=\sum\limits_{l=0}^{L}(-j)^l(2l+1)h_l^{(1)}(kr_{pq})P_l(\mathbf{\hat{k}}\cdot\mathbf{\hat{r}}_{pq}),
\end{equation} where $L=kD+2.9(kD)^{\nicefrac{1}{3}}$ is the truncation number of an infinite series and its physical meaning is the spatial bandwidth of the radiative system \cite{ref22}. Here, $D$ is the maximal size of the aperture; $h_l^{(1)}$ is the lth-order spherical Hankel function of the first kind\footnote{Note that in this paper, the definition of the term is $h_l^{(1)}=j_l(x)-jy_l(x)$, where $j_l(x)$ and $y_l(x)$ are the spherical Bessel and Neumann functions, respectively.}, and $P_l(x)$ is the $l$th-order Legendre polynomial. Substituting \eqref{eq3} and \eqref{eq4} into \eqref{eq1}, and changing the order of integrals based on the Fubini’s theorem, we then have

\begin{equation}
\label{eq6}
\begin{split}
    \mathbf{E(r)}& =-\frac{k\omega\mu}{16\pi^2}\int_{\boldsymbol{{\mathcal{A}}_t}}\left[\oint_{S_E}(\overline{\mathbf{I}}-\hat{\mathbf{k}}\hat{\mathbf{k}})e^{-jk\hat{\mathbf{k}}\cdot\mathbf{r}_{qs}}\cdot\alpha(\hat{\mathbf{k}}\cdot\hat{\mathbf{r}}_{pq})  \cdot e^{-jk\hat{\mathbf{k}}\cdot\mathbf{r}_{rp}}d\hat{\mathbf{k}}\right]\cdot\mathbf{J}(\mathbf{s})d\mathbf{s} \\
    &\overset{(a)}{=}-\frac{k\omega\mu}{16\pi^2}\oint_{S_E}(\overline{\mathbf{I}}-\hat{\mathbf{k}}\hat{\mathbf{k}})\alpha(\hat{\mathbf{k}}\cdot\hat{\mathbf{r}}_{pq})\mathcal{F}[\mathbf{J}(\mathbf{s})]e^{-jk\hat{\mathbf{k}}\cdot\mathbf{r}_{rp}}d\hat{\mathbf{k}} \\
    &\overset{(b)}{=}-\frac{k\omega\mu}{16\pi^2}\mathcal{F}^{-1}\left\{(\overline{\mathbf{I}}-\mathbf{\hat{k}}\mathbf{\hat{k}})\alpha(\hat{\mathbf{k}}\cdot\hat{\mathbf{r}}_{pq})\mathcal{F}[\mathbf{J}(\mathbf{s})]\right\}.
\end{split}
\end{equation} In (6a), we change the order of integral, and note that $\mathbf{r}_{qs}$ is only the function of the source position. In (6b), we change the direction of the position vector $\mathbf{r}_{pr}=-\mathbf{r}_{rp}$ , and note that $\mathbf{r}_{pr}$ is only the function of the field position. Equation \eqref{eq6}, also well-known as the diagonalized form of the fast multipole expansion \cite{ref51}, illustrates that the field-source interaction in free space can be decomposed into three distinct steps. Initially, the current on the transmitter is transformed into the wave-number domain through Fourier transform, which is referred to the aggregation process. Subsequently, the information of the source is translated between the centers $(q,\,p)$, which is the translation process. Finally, the inverse Fourier transform is performed on the wave-number domain to obtain the field on the receiver, which is known as the disaggregation process. The equation serves as the basis of FMM and multilevel fast multipole algorithm (MLFMA), which represent the milestone of computational electromagnetics around 2000s and enable fast and accurate full-wave simulations of complicated targets that had been impossible to handle. Furthermore, Eq. \eqref{eq6} yields the following three inspirations:

1) The Fourier plane spectrum theory remains applicable in the near-field scene. The currently rapid development of near-field theory necessitates the utilization of complex mathematical techniques to address quadratic nonlinear phase terms in channel models. Nevertheless, the Fourier theory continues to serve as a valuable tool due to its favorable properties, clear mathematical physics insights, and the benefit of accelerating numerical calculations with the assistance of fast Fourier transforms (FFT).

2)	 By employing FMM, the free-space propagator can be decomposed into three constituent parts, as illustrated in Fig. \ref{fig2}\subref{fig2a}. A pivotal aspect to consider is that the translator quantifies the contribution of each wavenumber, thus we define it as a spatial filter. The block diagram is analogous to that depicted in \cite{ref54}, but we use an underlying EM consistent approach and do not incorporate random terms.

3)	 An interesting observation is that the structure of Fig. \ref{fig2}\subref{fig2a} is in analogy with that of a conventional 4f system in optics, as illustrated in Fig. \ref{fig2}\subref{fig2c}. In this context, the received field is an image of the current function, whereas the spatial filter stands on the spectral plane.

\subsection{Finite Angular Spectral Bandwidth Characteristics}
In a numerical evaluation of the radiation integral according to \eqref{eq6}, the computational cost primarily dependents on the number of $\mathbf{\hat{k}}$ vectors sampled on the unit sphere ($0\leq\theta\leq \pi$ and $0\leq\varphi\leq 2\pi$). Considering the spatial bandwidth of the radiative system, the truncation number $L$ quantifies the angular resolution that depends on the maximal size of the aperture. In the conventional FMM, $\varphi$ is equidistantly sampled with $2L$ points, and $\theta$ is sampled with $L$ points with Gauss--Legendre principle. Consequently, the total number of $\mathbf{\hat{k}}$ vectors is approximately $2L^2$. The aggregation, translation, and disaggregation are exceedingly time-consuming for the radiation and scattering problems of large arrays and objects.

Further analysis of the translator reveals that only the $\mathbf{\hat{k}}$ vectors in the vicinity of the center direction $\mathbf{\hat{r}}_{pq}$ between the transmitter and receiver exhibit a significant contribution, which constitutes the bases of the ray propagation fast multipole algorithm (RPFMA) \cite{ref54,ref55}.  Consequently, the majority of plane-wave samples can be abandoned, as their magnitude falls below a given tolerable threshold. To fully leverage this idea, a window function is applied to the translator, thereby enhancing the contributions of the main directions while attenuating those of the minor. The following equation illustrates the utilization of a Tukey window function, specifically,
\begin{equation}
\label{eq7}
\alpha(\mathbf{\hat{k}}\cdot\mathbf{\hat{r}}_{pq})=\sum\limits_{l=0}^{L}(-j)^l(2l+1)h_l^{(1)}(kr_{pq})P_l(\mathbf{\hat{k}}\cdot\mathbf{\hat{r}}_{pq})w_l,
\end{equation} where the weight $w_l$ is one if $l\in[0,\,L/2]$ and has a cosine taper on the interval $(L/2,\,L]$ . The employment of the window function serves to accentuate the finite angular spectral bandwidth characteristics of the translator, while simultaneously yielding nearly identical results with a faster computational rate (refer to Fig. 1 in \cite{ref55}).

\begin{figure}[!h]
\centering
\subfloat[]{\includegraphics[width=0.45\textwidth]{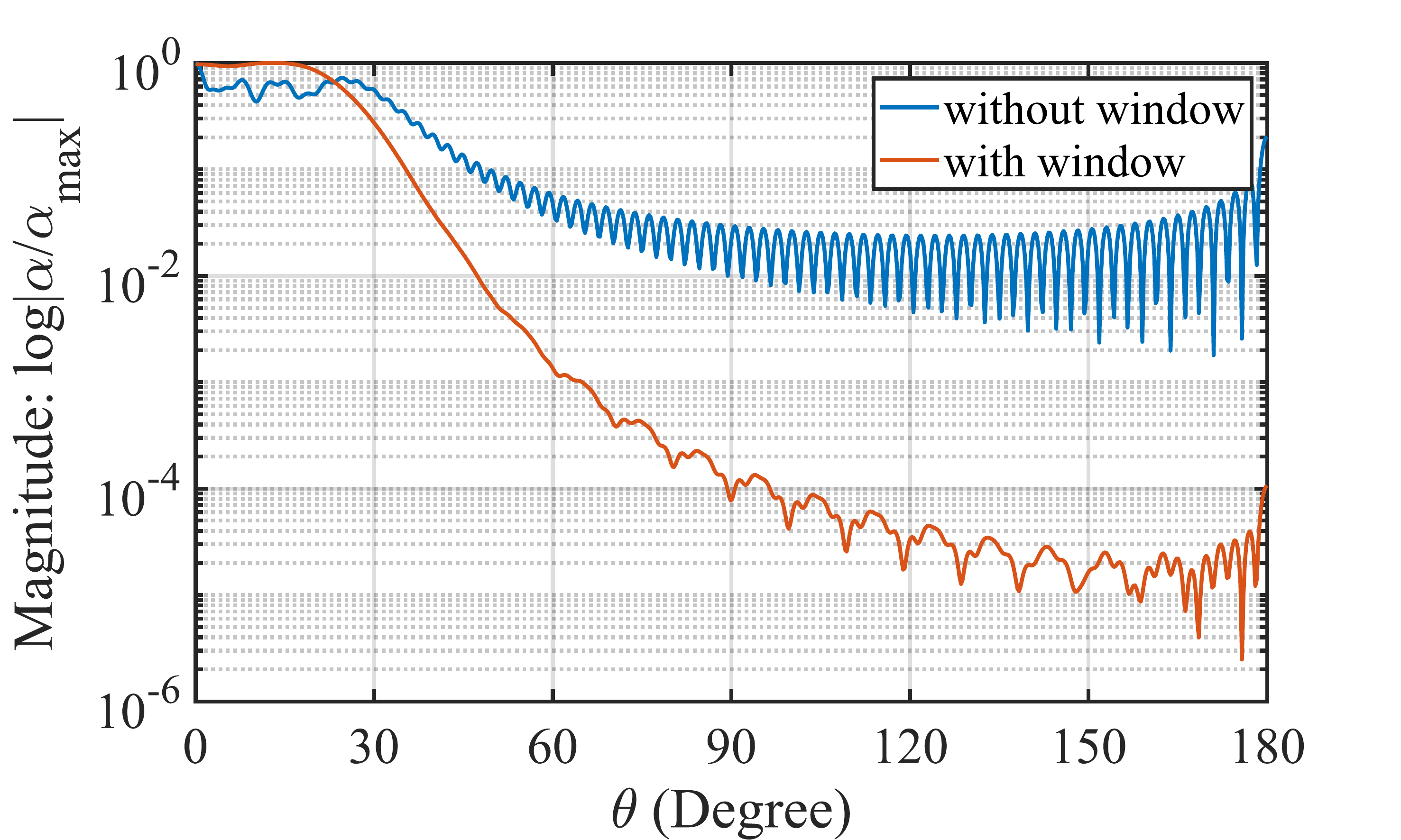}%
\label{fig3a}}
\hfil
\subfloat[]{\includegraphics[width=0.45\textwidth]{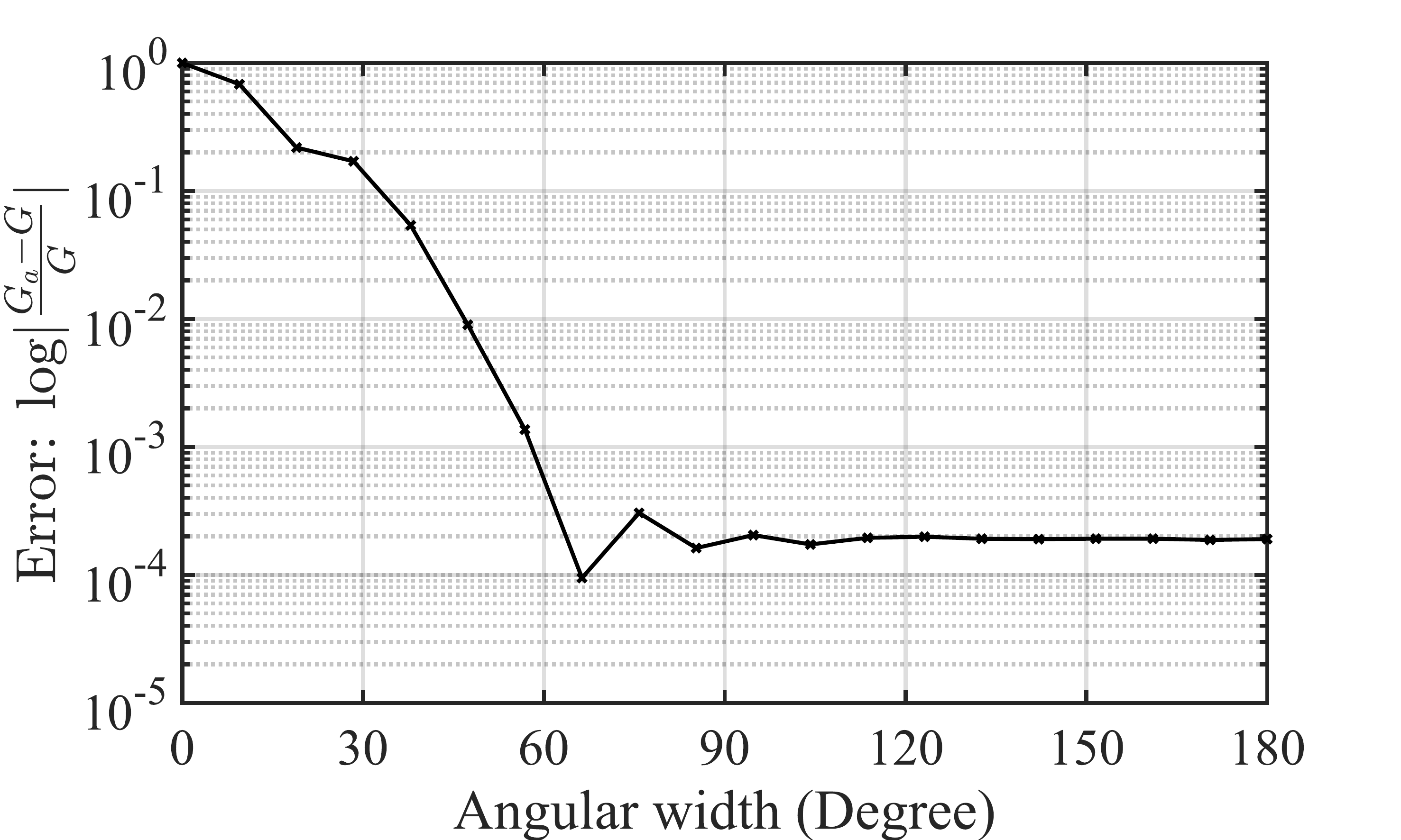}\label{fig3b}}%
\caption{Illustration for understanding the finite-bandwidth property of the translator and \eqref{eq4}. (a) The normalized translators with and without the window function. The aperture is $10\lambda$ and the distance between the transmitter and the receiver is $20\lambda$. The source position is $\mathbf{s}=(-5,\,1,\,1)\lambda$, the field is $\mathbf{r}=(-3.5,\,5,\,20)\lambda$, and their centers are $(0,\,0,\,0)\lambda$ and $(0,\,0,\,20)\lambda$, respectively. (b) The relative error as a function of the sampled angular width $\theta_e$. $G_a$ represents the integral on the conical surface $S_{\theta_e}$ defined by \eqref{eq4}, and $G$ is the value calculated by $e^{-jk|\mathbf{r}-\mathbf{s}|}/(4\pi|\mathbf{r}-\mathbf{s}|)$, which is accurate.}
\label{fig3}
\end{figure}

As an illustrative example, let us consider the following scenario where the transmitter aperture size is $10\lambda$, and the distance between the transmitter and the receiver is $20\lambda$. In this instance, the translator is merely a function of the angle between the vectors and the center direction $\mathbf{\hat{r}}_{pq}$, i.e., $\cos{\theta}=\mathbf{\hat{k}}\cdot\mathbf{\hat{r}}_{pq}$. Fig. \ref{fig3}\subref{fig3a} illustrates the normalized magnitude of the translator with and without the window functions. Obviously, the translator with the window function is smoother and decays faster, thereby discarding more $\mathbf{\hat{k}}$ vectors. Therefore, it is feasible to solely sample the $\mathbf{\hat{k}}$ vectors in the vicinity of the center direction $\mathbf{\hat{r}}_{pq}$, significantly reducing the computational cost. In other words, the integral of \eqref{eq4} can be performed on a conical surface $S_{\theta_e}$ of finite width $\theta_e$ with a tiny error. The physical meaning is that the translator is of finite spatial bandwidth, therefore, only part of the spatial spectrum in the source information shall be kept, similar to the approach in the time-frequency domain signal processing. As illustrated in Fig. \ref{fig3}\subref{fig3b}, the results confirm the above statement: as the angular width increases, the relative error gradually decreases and eventually becomes stable. Specifically, the abscissa is the angular width $\theta_e$, and the ordinate is the relative error $\log\left|(G_a-G)/G\right|$, in which $G_a$ represents the integral on the conical surface $S_{\theta_e}$ according to the right of \eqref{eq4} and it is the function of angular width, and $G$ is the value calculated by  $e^{-jk|\mathbf{r}-\mathbf{s}|}/(4\pi\mid\mathbf{r}-\mathbf{s}\mid)$, which is accurate and constant.
\begin{figure}[!ht]
    \centering
    \includegraphics[width=0.5\textwidth]{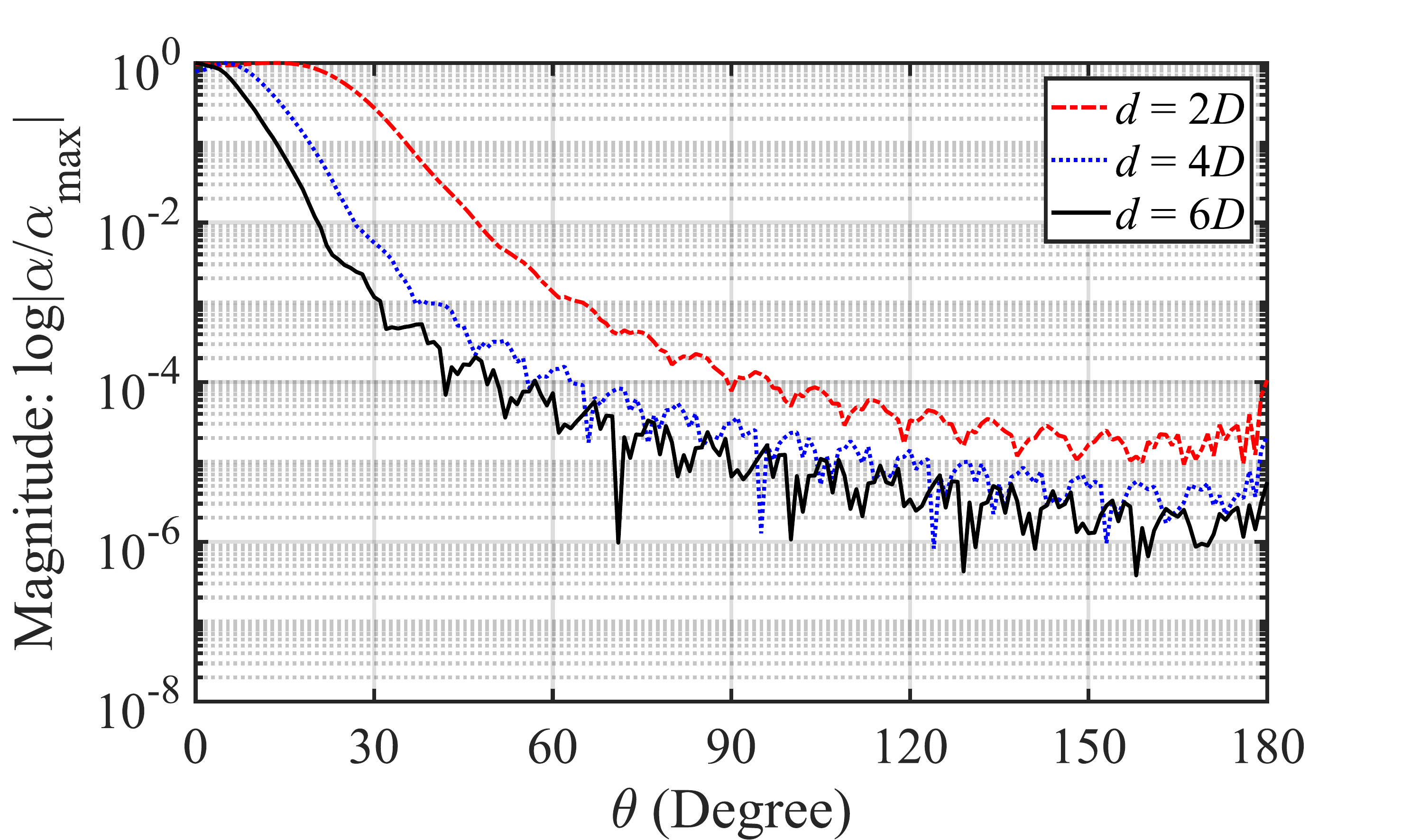}
    \caption{Illustration for understanding the relationship between the angular bandwidth and the distance between the transmitter and receiver. The configurations of the source and field regions are the same, we just change the distance.}
    \label{fig4}
\end{figure}
In this section, we present two spatial bandwidths that appear to be distinct, namely the source bandwidth $L$ and the channel bandwidth $\theta_e$. The former quantifies the angular resolution of the transmitter and depends on the transmitting aperture, thereby determining the spectra of the source information. The latter, also known as the finite angular bandwidth, quantifies the contributions of each spectrum of the source information and depends on the distance between the transmitter and the receiver. When the distance increases, the angular bandwidth reduces, as shown in Fig. \ref{fig4}, and thus the effective spectra of the source information decrease. We indicate that the effective spectra, which can be seen as the spatial DoF of the system, are determined by the aperture of the transmitter and the receiver, and their distance.

\section{System Capacity}
\subsection{Design of Precoder and Combiner}
As previously stated, to design the precoder $\mathbf{Q}[\cdot]$ and combiner $\mathbf{C}[\cdot]$, it is essential to consider the design of the current distribution function that achieves the maximal power transmission between the transmitter and receiver. For simplicity, we focus on a single polarization scenario.

The optimization problem is formulated as
\begin{equation}
\label{eq8}
\begin{split}
&\max_{J(\mathbf{s})}\quad\frac1{2\eta}\int_{\boldsymbol{{\mathcal{A}_r}}}\mid E(\mathbf{r})\mid^2d\mathbf{r} \\
    &\text{ s. t. }\quad\int_{\boldsymbol{{\mathcal{A}_r}}}\eta\mid J(\mathbf{s})\mid^2d\mathbf{s}=P_t,
\end{split}
\end{equation} where $\eta$ and $P_t$ represent the free space impedance and the transmit power, respectively. After some mathematical operations (see Appendix \ref{appA} for more details), the problem \eqref{eq8} is transformed into the solution of a second-kind homogeneous Fredholm integral equation, namely,
\begin{equation}
\label{eq9}
\beta\varphi(\mathbf{s})=\int_{\boldsymbol{\mathcal{A}_t}} M(\mathbf{s,\,s'})\varphi(\mathbf{s'}) d\mathbf{s'},
\end{equation} where $\beta$ and $\varphi(\mathbf{s})$ are the eigenvalue and eigenfunction. Specifically, $\varphi(\mathbf{s})$ represents the mode current function supporting the maximal power transmission, and $\beta$ quantifies the power level of the field on the receiver launched by the above current function. The integral kernel $M$ is expressed as
\begin{equation}
\label{eq10}
M(\mathbf{s,\,s'})=\int_{\boldsymbol{\mathcal{A}_r}}H(\mathbf{r,\,s})\bar{H}(\mathbf{r,\,s'})d\mathbf{r},
\end{equation} and $H(\mathbf{r,\,s})$ is
\begin{equation}
\label{eq11}
    H(\mathbf{r},\,\mathbf{s})=-\frac{k\omega\mu}{16\pi^2}\int_{S_{\theta_e}}e^{-jk\hat{\mathbf{k}}\cdot\hat{\mathbf{r}}_{qs}}\cdot\alpha(\hat{\mathbf{k}}\cdot\hat{\mathbf{r}}_{pq})\cdot e^{-jk\hat{\mathbf{k}}\cdot\hat{\mathbf{r}}_{rp}}d\hat{\mathbf{k}}, 
\end{equation} where $M$ is Hermitian since $M(\mathbf{s,\,s'})=\bar{M}\mathbf{(s',\,s)}$. In this context, the solution of \eqref{eq9} generates a sequence of eigenvalue and eigenfunction pairs $(\beta_n,\,\varphi_n(\mathbf{s}))$, which satisfy the following three favorable properties (see Appendix \ref{appB} for more details):

1) The eigenvalue $\beta_n$ is real. More importantly, since the integral kernel $M$ is compact and bounded on the transmitter $\boldsymbol{\mathcal{A}_t}$, the first $N$ eigenvalues are almost equal $\beta_1\approx\beta_2\approx\cdots\approx\beta_N$, and decrease sharply when $n>N$ \cite{ref45}.

2) The sets of eigenfunctions $\{\varphi_n(\mathbf{s})\}_{n=1}^{\infty}$ are orthogonal on $\boldsymbol{\mathcal{A}_t}$ and complete in $\mathcal{L}^2(\boldsymbol{\mathcal{A}_r})$ space, specifically,
\begin{equation}
\label{eq12}
\int_{\boldsymbol{\mathcal{A}_t}}\varphi_m(\mathbf{s})\bar{\varphi}_n(\mathbf{s})d\mathbf{s}=\delta_{mn}.
\end{equation}

3) The electric field $\psi_n(\mathbf{r})$ induced by the mode current $\varphi_n(\mathbf{s})$ on the receiver is expressed as
\begin{equation}
\label{eq13}
\psi_n(\mathbf{r})=\int_{\boldsymbol{\mathcal{A}_t}}H(\mathbf{r,\,s})\cdot\varphi(\mathbf{s})d\mathbf{s},\quad\mathbf{r}\in\boldsymbol{\mathcal{A}_r},
\end{equation} and it is biorthogonal, which is orthogonal on the whole free space $\boldsymbol{\mathcal{A}_{sum}}$ and orthogonal on the receiver $\boldsymbol{\mathcal{A}_{r}}$, namely,
\begin{equation}
\label{eq14}
    \begin{split}
        &\int_{\boldsymbol{\mathcal{A}_{sum}}}\psi_m(\mathbf{r})\bar{\psi}_n(\mathbf{r})d\mathbf{r}=\delta_{mn},\quad \text{orthonormal}\\
        &\int_{\boldsymbol{\mathcal{A}_r}}\psi_m(\mathbf{r})\bar{\psi}_n(\mathbf{r})d\mathbf{r}=\beta_m\delta_{mn},\quad \text{orthogonal}.
    \end{split}
\end{equation}

According to the above properties, arbitrary current on the transmitting aperture can be represented using a linear combination of the orthogonal functions, specifically,
\begin{equation}
\label{eq15}
    J(\mathbf{s})=\sum\limits_{n=1}^{\infty} x_n\varphi_n(\mathbf{s}),\quad\mathbf{s}\in\boldsymbol{\mathcal{A}_s},
\end{equation} where $x_n$ is the constellation symbol and follows a certain random distribution. To represent the arbitrary field on the receiver, we define the orthogonal functions, namely,
\begin{equation}
\label{eq16}
    \chi_n(\mathbf{r})=\frac{1}{\sqrt{\beta_n}}\psi_n(\mathbf{r}),\quad\mathbf{r}\in\boldsymbol{\mathcal{A}_r}.
\end{equation}

At this stage, we have obtained the precoder and the combiner, as well as their coupling strengths, which are $\mathbf{Q}[\cdot]=\{\varphi_n(\mathbf{s})\}_{n=1}^{\infty}$, $\mathbf{C}[\cdot]=\{\chi_n(\mathbf{r})\}_{n=1}^{\infty}$, and $\{\beta_n\}_{n=1}^{\infty}$, respectively. Since $\chi_n(\mathbf{r})$ is also complete on $\boldsymbol{\mathcal{A}_r}$, the field and interference can be represented as
\begin{equation}
\label{eq17}
    E(\mathbf{r})=\sum_{n=1}^\infty e_n\chi_n(\mathbf{r}),\quad N(\mathbf{r})=\sum_{n=1}^\infty\sigma_n\chi_n(\mathbf{r}),
\end{equation} where $e_n$ represents the received symbol, and $\sigma_n$ is the noise of the $n$th subchannels. In this analysis, we consider the complex additive white Gaussian noise, which is assumed to be independent of the channel. Thus, we employ $\sigma$ to represent the noise of each subchannel, i.e., $\sigma=\sigma_n$. Furthermore, the received field is $E_{noise}(\mathbf{r})=E(\mathbf{r})+N(\mathbf{r})$ in the presence of noise. Substituting \eqref{eq15} into \eqref{eq1}, and considering \eqref{eq13}, \eqref{eq16}, and \eqref{eq17}, the explicit expression of the received field is (see Appendix \ref{appC} for more details),
\begin{equation}
\label{eq18}
    E_{noise}(\mathbf{r})=\sum_{n=1}^\infty\left[\sqrt{\beta_n}x_n+\sigma\right]\chi_n(\mathbf{r}).
\end{equation} Equation \eqref{eq18} indicates that the radiating system has infinity independent subchannels between the transmitter and the receiver. Given the attenuation properties of $\beta_n$, only the first $N$ items have significant contributions, thus we can truncate \eqref{eq18} as
\begin{equation}
\label{eq19}
    E_{noise}(\mathbf{r})=\sum_{n=1}^N\left[\sqrt{\beta_n}x_n+\sigma\right]\chi_n(\mathbf{r}).
\end{equation}

\subsection{The Effective Spatial DoF}
The closed-form solution of spatial DoF has been discussed from many different perspectives, including the diffractive theory, the signal space, and the analytical method, which is formulated as,
\begin{equation}
\label{eq20}
    N=\frac{\left|\mathcal{A}_t|\cdot|\mathcal{A}_r\right|}{\lambda^2d^2},
\end{equation} where $|\boldsymbol{\mathcal{A}_t}|$, $\vert\boldsymbol{\mathcal{A}_r}\vert$, and $d$ represent represent the areas of the transmitter and receiver, and the distance between them, respectively. The above equation stated that the spatial DoF of the system, is determined by the geometry of the transmitter and the receiver, specifically the apertures and their distance. We here provide a signal recovery method for deriving the effective spatial DoF. The newly defined DoF allow us to concentrate on the  
relationship between the spatial DoF, the noise, and the transmitting power from a signal recovery perspective.

Since $E(\mathbf{r})$ contains the source information without the noise and the receiver can only obtain the field $E_{noise}(\mathbf{r})$ with noise, we let the mean square error minimal by approximating $E_{noise}(\mathbf{r})$ to $E(\mathbf{r})$, with the number of truncated terms $N_E$ serving as the effective spatial DoF. The $N_E$ is more accurate than the estimated value $N$ defined by \eqref{eq20}. In this context, the mean square error is

\begin{equation}
    \label{eq21}
    \begin{split}
        \varepsilon_{N_{E}}^{2}&=\mathbb{E}\left[\int_{{\mathcal{A}_r}}|E_{noise}(\mathbf{r})-E(\mathbf{r})|^2 d\mathbf{r}\right]  \\   
&=\mathbb{E}\left[\int_{{\mathcal{A}_r}}\left|\sum_{n=1}^{N_E}\left(\sqrt{\beta_n}x_n+\sigma\right)\chi_n(\mathbf{r})-\sum_{n=1}^\infty\sqrt{\beta_n}x_n\chi_n(\mathbf{r})\right|^2d\mathbf{r}\right]  \\
 & \overset{(a)}{=}\sum_{j=N_E+1}^\infty\beta_n\mathbb{E}\left[\mid x_n\mid^2\right]+\sum_{j=1}^{N_E}\mathbb{E}\left[\mid\sigma\mid^2\right]  \\
&\leq\mathbb{E}\left[\mid x_{N_E+1}\mid^2\right]\sum_{i=1}^\infty\beta_n-\sum_{i=1}^{N_E}\left(\beta_n\mathbb{E}\left[\mid x_{N_E+1}\mid^2\right]-\mathbb{E}\left[\mid\sigma\mid^2\right]\right).
    \end{split}
\end{equation}

In (21a), we use the orthogonality of the function $\psi_n{\mathbf{r}}$. The inequality holds since we use the assumption, that is, $\mathbb{E}{\left[|x_{N_{E}+1}|^2\right]}\geq\mathbb{E}{\left[|x_{N_{E}+2}|^2\right]}\geq\cdots$. It is worth noting that, to achieve a higher capacity, the transmitter tends to allocate a greater proportion of power to the subchannels with higher eigenvalues, which means that the allocated power of the $n$th subchannel $\mathbb{E}{\left[|x_n|^2\right]}$ is positively correlated with the eigenvalue $\beta_n$. Notice that the first term in \eqref{eq21} is constant, thus the condition to make $\varepsilon_{N_E}^2$ minimal is to ensure the second term maximal. Since the eigenvalue $\beta_n$ decreases as the mode index n increases, the second term can reach the maximum when the following conditions satisfies
\begin{equation}
\label{eq22}
    \beta_{N_E}\mathbb{E}\left[\mid x_{N_E+1}\mid^2\right]-\mathbb{E}\left[\mid\sigma\mid^2\right]=0.
\end{equation} 
Therefore, the effective spatial DoF $\beta_{N_E}$ is the index number of the eigenvalue that ensures $\beta_{N_E}=\mathbb{E}\left[\mid\sigma\mid^2\right]\mathbb{E}\left[\mid x_{N_E+1}\mid^2\right]$. Equation \eqref{eq22} states that the effective DoF takes into account the channel feature characterized by the eigenvalues and the signal noise ratio (SNR) of the radiating system. Since the distribution of the eigenvalues is deterministic for a given configuration of the transmitter and the receiver, the number of the effective DoF also dependents on the SNR of the system. Nevertheless, we can observe that the distribution of the eigenvalues shows two features: 1) the first $N$ eigenvalues defined by \eqref{eq20} are nearly identical; 2) the magnitude of the eigenvalue declines rapidly with mode number when $n \geq N$.  Consequently, an increase in SNR at a considerable expense may only yield a few supplementary and inconsequential spatial degrees of freedom. Therefore, although the spatial degree of freedom $N$, as defined by \eqref{eq20}, solely considers the geometry of transmitter and the receiver, it remains a compromise approximation.

\subsection{The EM Channel Capacity Formula}
We here consider the first $N_E$ effective subchannels, namely,
\begin{equation}
\label{eq23}
    \hat{x}_n=\sqrt{\beta_n}x_n+\sigma,\quad n=1,\,2,\,\cdots,\,N_E
\end{equation} subject to an average total current power constraint of $P_t$:
\begin{equation}
\label{eq24}
    \begin{aligned}
\mathbb{E}\left[\int_{\mathcal{A}_{t}}\eta | J(\mathbf{s}) |^{2}d\mathbf{s}\right]& =\mathbb{E}\left[\int_{\mathcal{A}_{\mathbf{r}}}\left|\sum_{n=1}^{\infty}x_{n}\varphi(\mathbf{s})\right|^{2}d\mathbf{s}\right] \\
&=\sum_{n=1}^\infty\mathbb{E}\left[\mid x_n\mid^2\right]=P_t.
\end{aligned}
\end{equation}

The capacity in bits per channel use is
\begin{equation}
\label{eq25}
    \begin{aligned}C_{N_{E}}&=\max_{\sum\limits_{n=1}^{N_E}\mathbb{E}\left[|x_n|^2\right]\leq P_t}I(\mathbf{x};\,\hat{\mathbf{x}})\\
    &\leq\sum_{n=1}^{N_E}\log(1+\frac{\alpha_nP_n}\mu),
\end{aligned}
\end{equation} where $\mathbf{x}=(x_1,\,x_2,\,\cdots,\,x_{N_E})^T$ is the vector containing the transmitted symbols, $\mathbf{\hat{x}}=(\hat{x}_1,\,\hat{x}_2,\,\\\cdots,\,\hat{x}_{N_E})^T$ is the vector containing the received symbols, $P_n=\mathbb{E}\left[|x_n|^2\right]$ is the allocated power of the nth subchannel, and $\mu=\mathbb{E}\left[\sigma^2\right]$  is the noise power of the nth subchannel. The equality holds in \eqref{eq25} when the signal $x_n$ and the noise $\sigma$ follow the complex Gaussian distribution, namely, $\mathcal{CN}(0,\,P_n)$ and $\mathcal{CN}(0,\,\mu)$, respectively \cite{ref56}. The maximization of the capacity $C_{N_E}$ can be reduced to a power allocation problem, which is resolved using the water-filling algorithm. The closed form of the allocated power of the nth subchannel is
\begin{equation}
\label{eq26}
    P_n=\frac{P_t}M+\frac1M\sum_{i=1}^M\frac{\sigma^2}{\beta_i}-\frac{\sigma^2}{\beta_n},\quad n=1,\,2,\,\cdots,\,M
\end{equation} where $M$ is the number of subchannels that obtain the allocated power, and thus $M<N_E$. For the remaining $N_E-M$ subchannels, the transmitter does not allocate power to them. We then can obtain the optimal capacity by substituting \eqref{eq26} into \eqref{eq25}, however, the expression seems complicated. 

Next, we give a concise and sub-optimal form. Considering the property of the first $N$ eigenvalues satisfy $\beta_1\approx\beta_2\approx\cdots\approx\beta_N\approx\beta$, the transmitter allocates the power of each subchannel to the same value $P_n=P_t/N$. Hence the capacity in bits per message is written as
\begin{equation}
\label{eq27}
    C_N=N\log(1+\frac{\beta P_t}{N\mu}),\quad\mathrm{bits}\cdot\mathrm{s}^{-1}\cdot\mathrm{Hz}^{-1}
\end{equation} where $N$ is the spatial DoF calculated by \eqref{eq20}. Furthermore, in subsequent numerical simulations, we compare the sub-optimal capacity calculated by \eqref{eq27} with the optimal case using the water-filling algorithm (i.e., \eqref{eq25}), which shows good consistence in the low and middle SNR regimes.
\section{Numerical Experiments}
\subsection{Simulation Configuration}
\noindent {1) \textit{Simulation Scenes}}

We consider a 3-D line-of-sight wireless communication scenario to illustrate our method and theory. Specifically, we employ a numerical method to solve the integral \eqref{eq9}, thereby obtaining explicit forms of the precoder and the combiner in the spatial domain (i.e., the eigenvalues and eigenfunctions). Once the complete channel information is obtained, we undertake a detailed analysis of the system capacity bound, taking into full consideration of the relationship among the spatial DoF, noise, and transmitted power. The transmitter and receiver are deployed on the xy-plane with their centers located at $(0,\,0,\,0)$ and $(0,\,0,\,d)$, respectively, namely,
\begin{equation}
\label{eq28}
    \begin{aligned}
\boldsymbol{\mathcal{A}_{t}}& =\left\{\left(s_{x},\,s_{y},\,s_{z}\right)\bigg{\vert}|s_{x}|\leq\frac{D_{tx}}{2}, \mid s_{y}\mid\leq\frac{D_{ty}}{2}, s_{z} =0\right\} \,\text{and} \\
\boldsymbol{\mathcal{A}_{r}}& =\left\{\left(r_x, r_y, r_z\right)\bigg{\vert}|r_x|\leq\frac{D_{rx}}{2},\,\mid r_y\mid\leq\frac{D_{ry}}{2},\,r_z =d\right\}, 
\end{aligned}
\end{equation} where the transmitter and the receiver have a square shape, $D_{tx}=D_{ty}=10\lambda$, $D_{rx}=D_{ry}=8\lambda$, and $d$is set to the sum of the aperture, namely, $d=\sqrt{2}(D_{tx}+D_{ty})=25.5\lambda$. ~\\
\noindent {2) \textit{Simulation Scenes}}

Once the geometry of the transmitter and the receiver has been defined, the classic Galerkin method is employed to resolve problem \eqref{eq9}, thereby enabling the precoder and combiner, along with their respective coupling strengths, to be obtained. In the Galerkin method, the basis is the Legendre function and the maximal order is 36. The details of the numerical method are shown in Appendix \ref{appE}. The truncation number of the translator formulated in \eqref{eq7} is about $L=93$. The sampled angular width is $\theta_e=60^{\circ}$, which means that the integral of \eqref{eq4} performs on the conical surface $S_{\theta_e}=\left\{(\theta,\,\varphi)\big|0^{\circ}\leq\theta\leq60^{\circ},\,0^{\circ}\leq\varphi\leq360^{\circ}\right\}$. Furthermore, the sampling number of integral $\int_{\boldsymbol{\mathcal{A}_t}}d\mathbf{s}$ and $\int_{\boldsymbol{\mathcal{A}_r}}d\mathbf{r}$ is set to 512, and the Gauss-Legendre knots and weights are utilized for numerical integral. The eigenvalue $\beta_n$ is normalized to one to observe its magnitude variation with respect to the channel mode index $n$. The maximal transmit power is set to $P_t=1$ W, and the noise power is $\sigma^2=10^{-SNR/10}$ W, which is changed according to the SNR (dB). 

\subsection{Results}
\begin{figure}[!ht]
    \centering
    \includegraphics[width=0.55\textwidth]{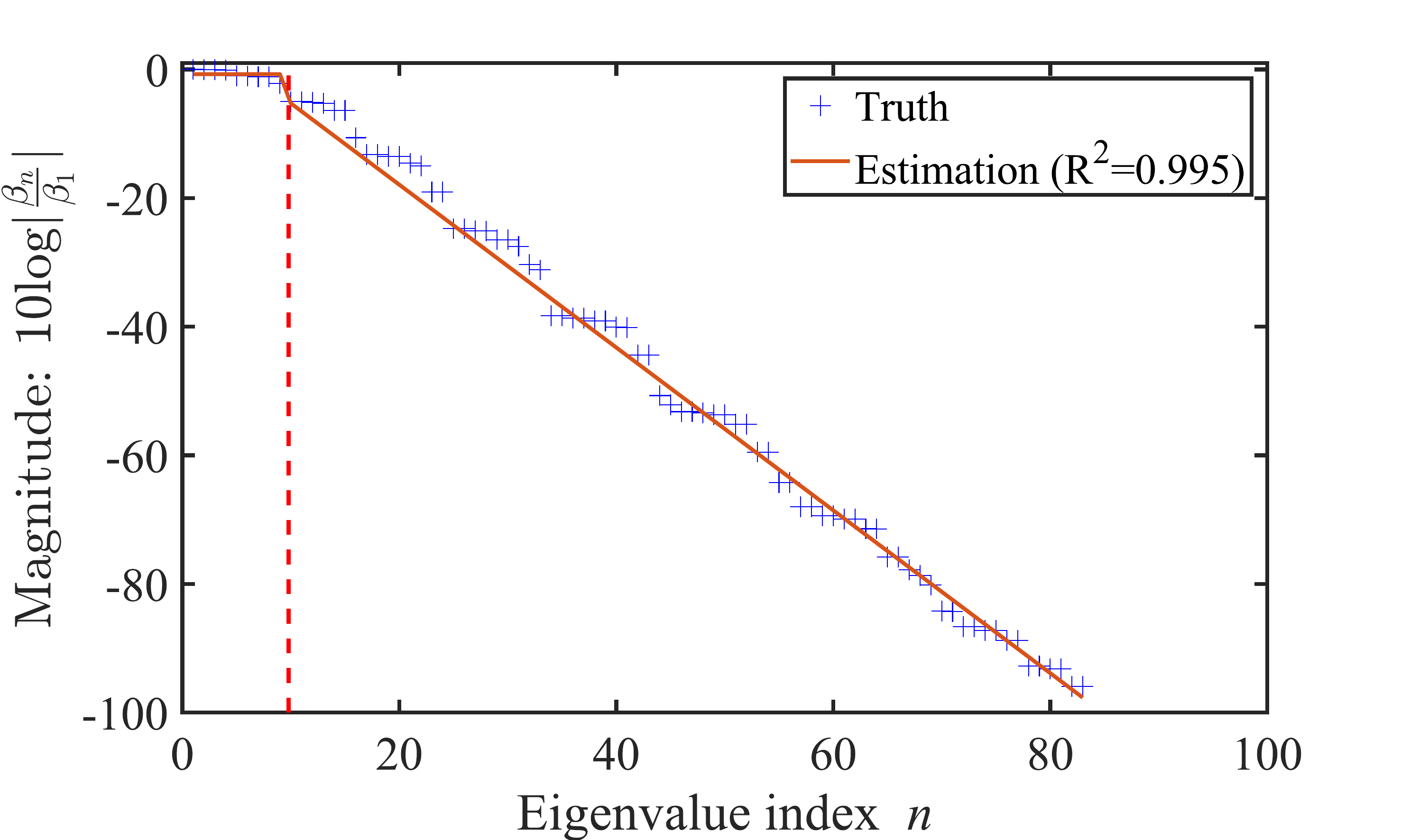}
    \caption{Illustration of the relative magnitudes of the eigenvalues $\beta_n$ with respect to the mode index. The abscissa of the vertical red dashed line is the value of the spatial DoF $N$ calculated by \eqref{eq20}, and the blue “+” marker represents the eigenvalues. The red solid line represents a piecewise function used for approximating the calculated eigenvalues.}
    \label{fig5}
\end{figure}

To validate the property of the eigenvalue $\beta_n$, Fig. \ref{fig5} illustrates the variation of the relative magnitude of the eigenvalues with respect to the mode index. The results exhibit two notable behaviors: 1) the relative magnitudes of the first N channels are nearly constants and bigger than -3 dB; and 2) the magnitudes apparently experience exponential decrease when the mode index exceeds $N$. We remark that the eigenvalue represents the coupling strength of the corresponding sub- channel and thus the results indicate that only the first $N$ channels are dominant. We employ a piecewise function to emulate the two primary characteristics of the eigenvalues, specifically,
\begin{equation}
\label{eq29}
    \beta_n=\begin{cases}\quad\frac{1}{N}\sum\limits_{i=1}^N\beta_n,&n\leq N\\\beta_{N+1}10^{-c(n-N-1)},&n\geq N+1.\end{cases}
\end{equation} In \eqref{eq29}, the eigenvalue is set to the average of the first $N$ terms if $n\leq N$. When $n$ exceeds $N$, its magnitude decreases exponentially with the mode index $n$, and the decay rate $c=0.127$ is obtained by a linear regression algorithm. It is observed that the estimation function is in good agreement with the eigenvalues obtained from the numerical method, with the correlation coefficients $R^2=0.995$.

\begin{figure*}[!ht]
\centering
\subfloat[]{\includegraphics[width=0.23\textwidth,height=7cm]{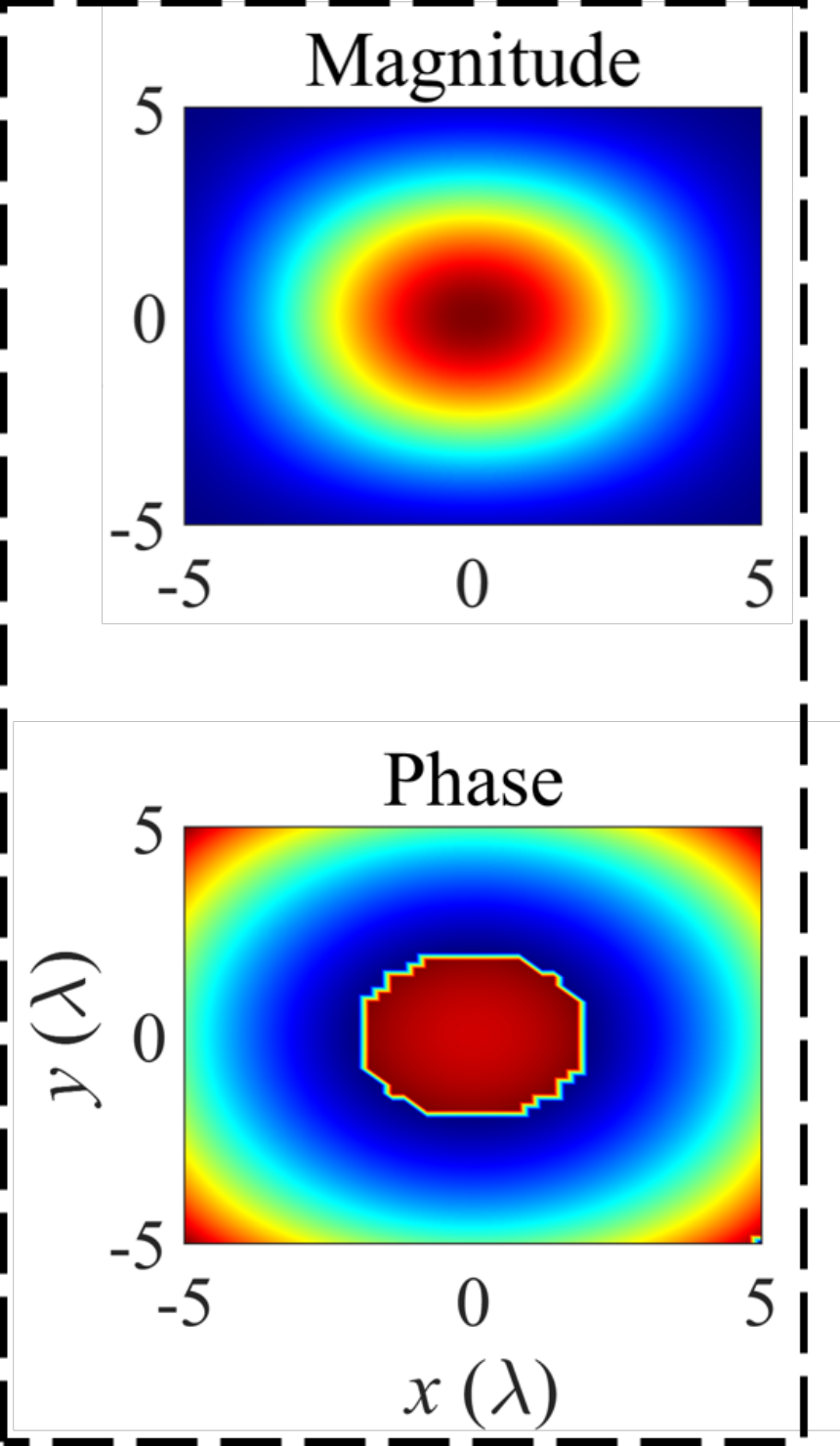}\label{fig6a}}%
\hfil
\subfloat[]{\includegraphics[width=0.23\textwidth,height=7cm]{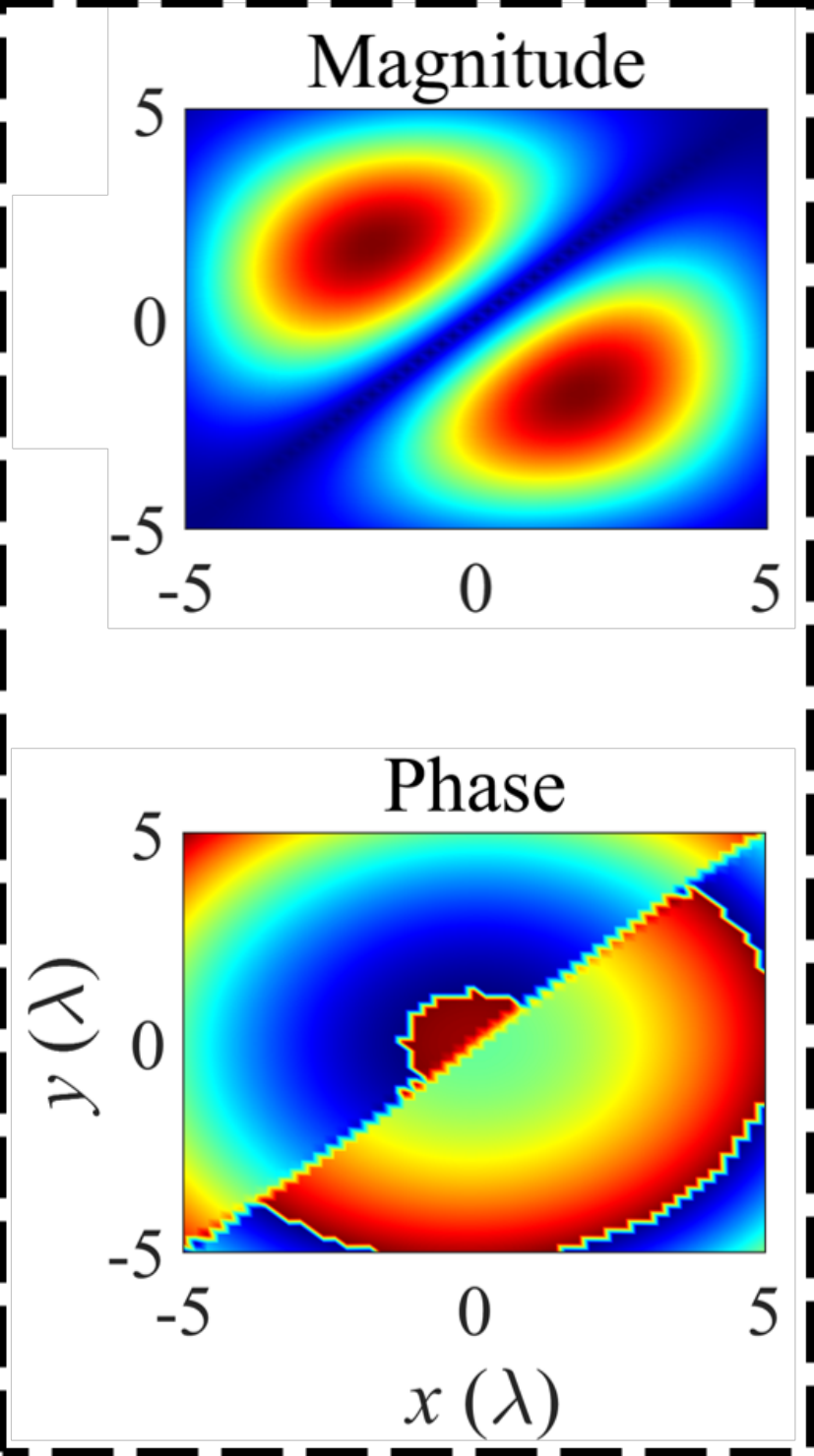}\label{fig6b}}%
\hfil
\subfloat[]{\includegraphics[width=0.23\textwidth,height=7cm]{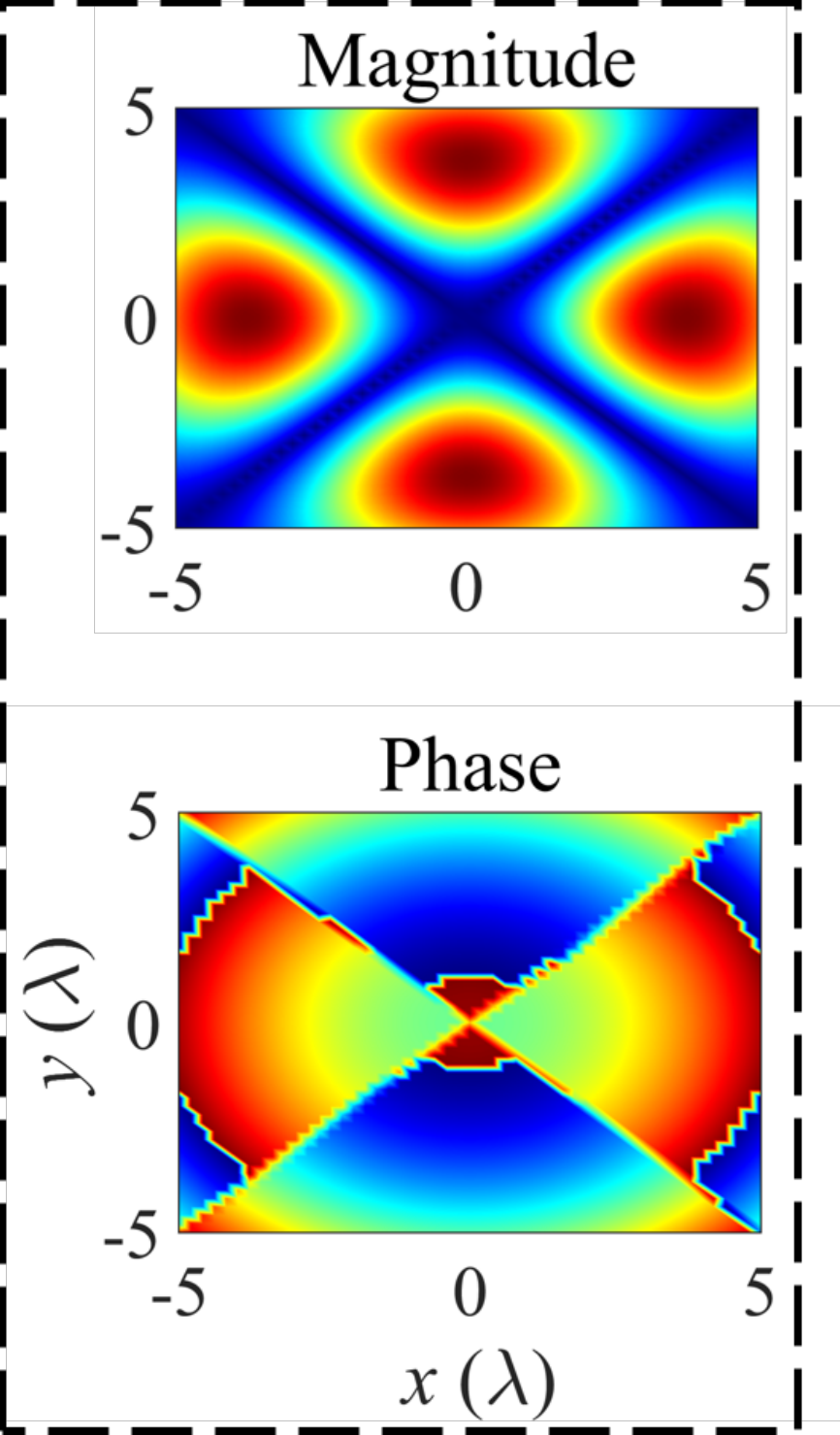}\label{fig6c}}%
\hfil
\begin{minipage}[b]{0.25\textwidth}
\centering
\subfloat[]{\includegraphics[width=1\textwidth,height=3.2cm]{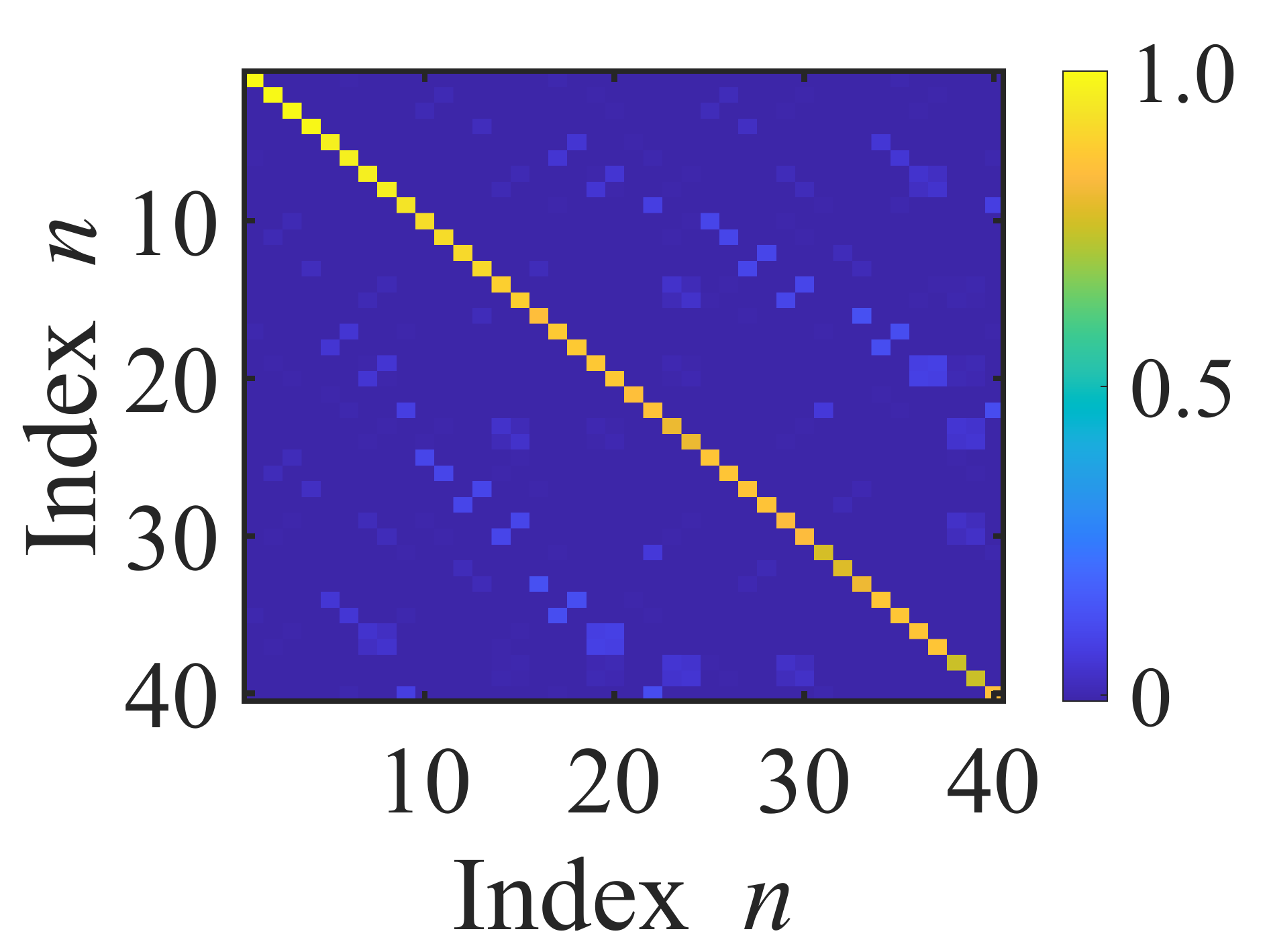}\label{fig6d}}
\vfill
\subfloat[]{\includegraphics[width=1\textwidth,height=3.2cm]{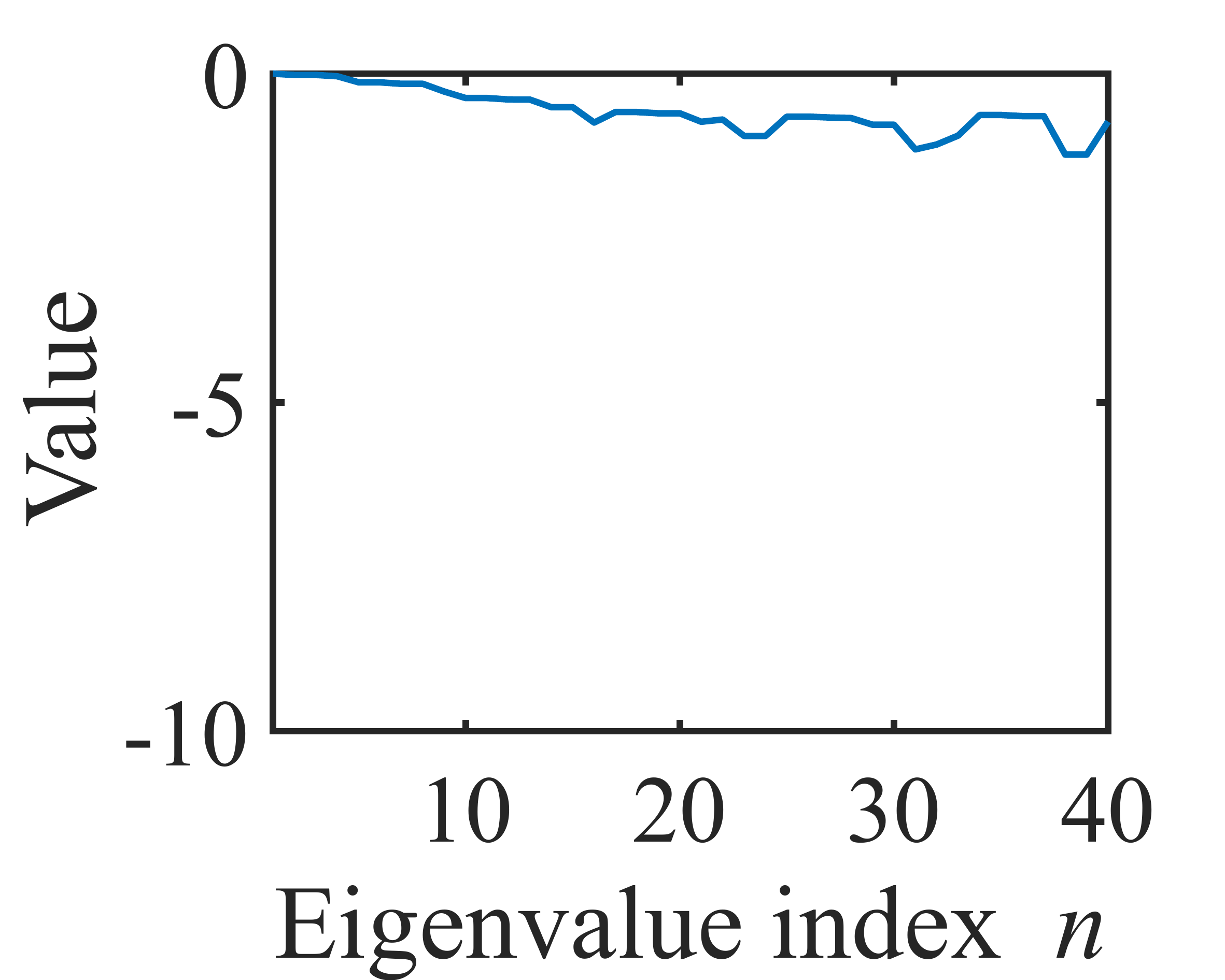}\label{fig6e}}
\end{minipage}
\caption{Illustration of the selected mode current distributions. (a) The magnitude and phase distributions of Mode 1. (b) The magnitude and phase distributions of Mode 3. (c) The magnitude and phase distributions of Mode 5. (d) The correlation matrix of the first 40 mode currents, which indicates the orthogonality of the mode currents. (e) The values on the diagonal of the correlation matrix for illustrating the normalized power of the nth mode current.}
\label{fig6}
\end{figure*}

\begin{figure*}[!ht]
\centering
\subfloat[]{\includegraphics[width=0.23\textwidth,height=7cm]{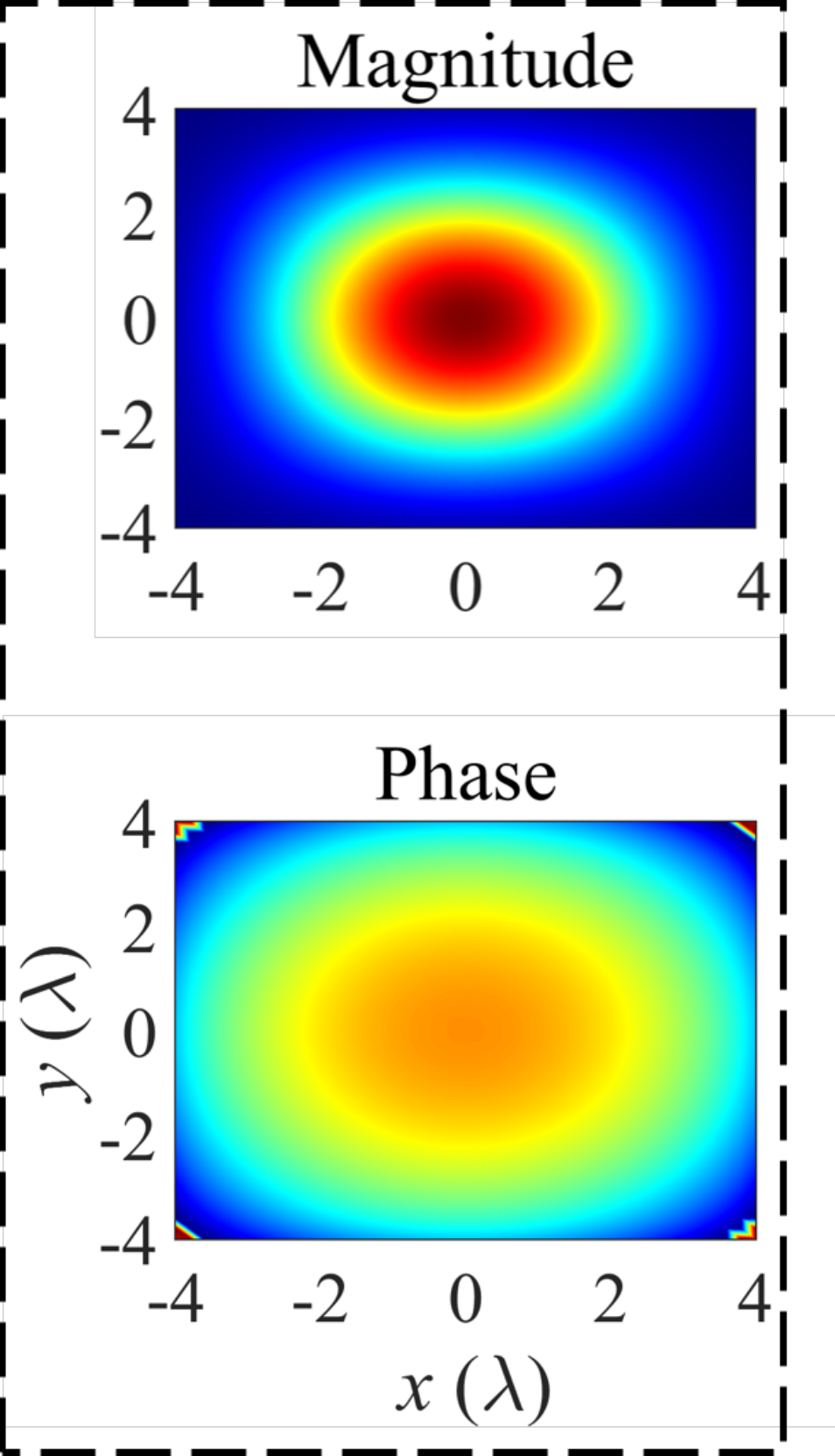}\label{fig7a}}%
\hfil
\subfloat[]{\includegraphics[width=0.23\textwidth,height=7cm]{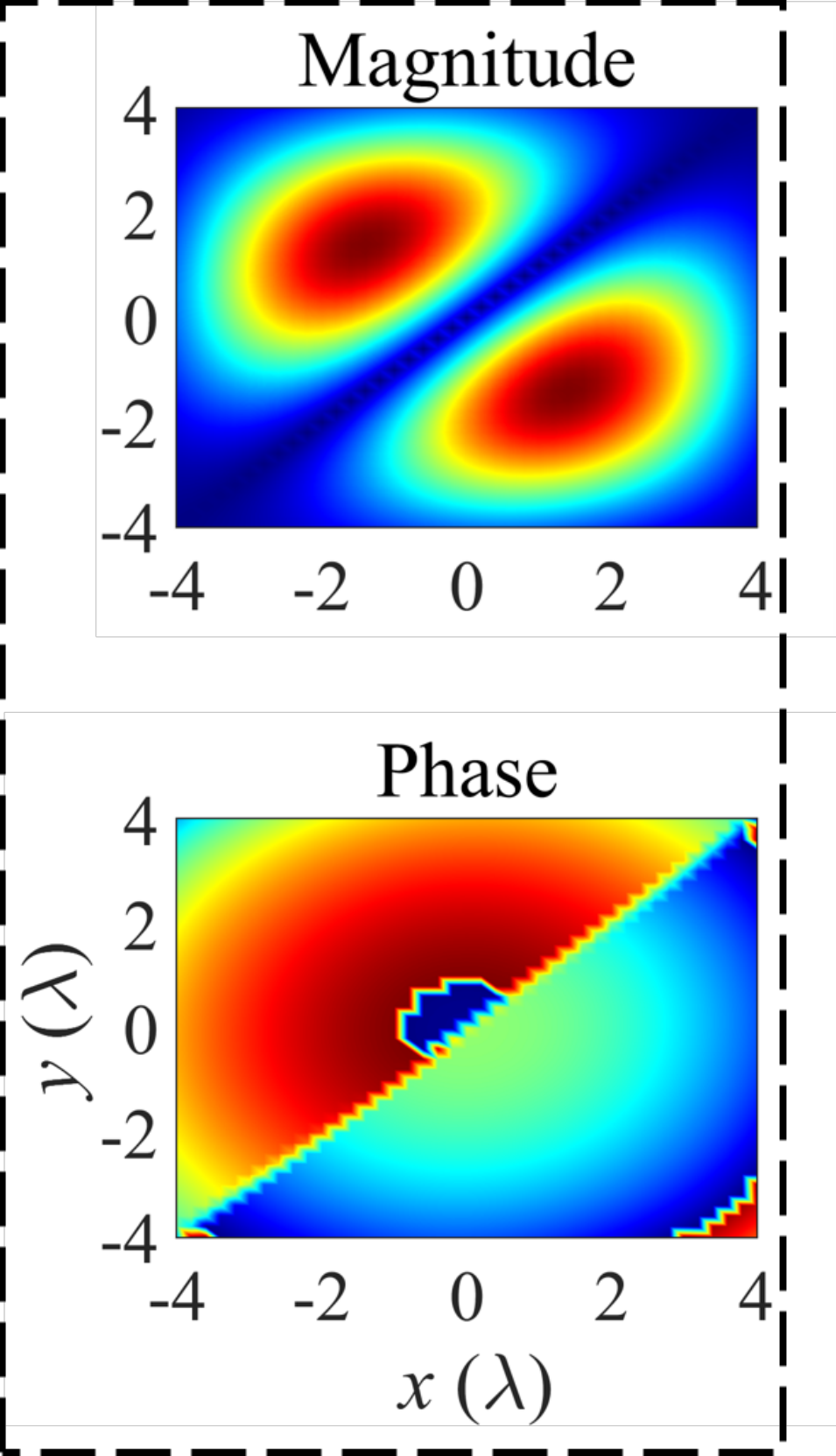}\label{fig7b}}%
\hfil
\subfloat[]{\includegraphics[width=0.23\textwidth,height=7cm]{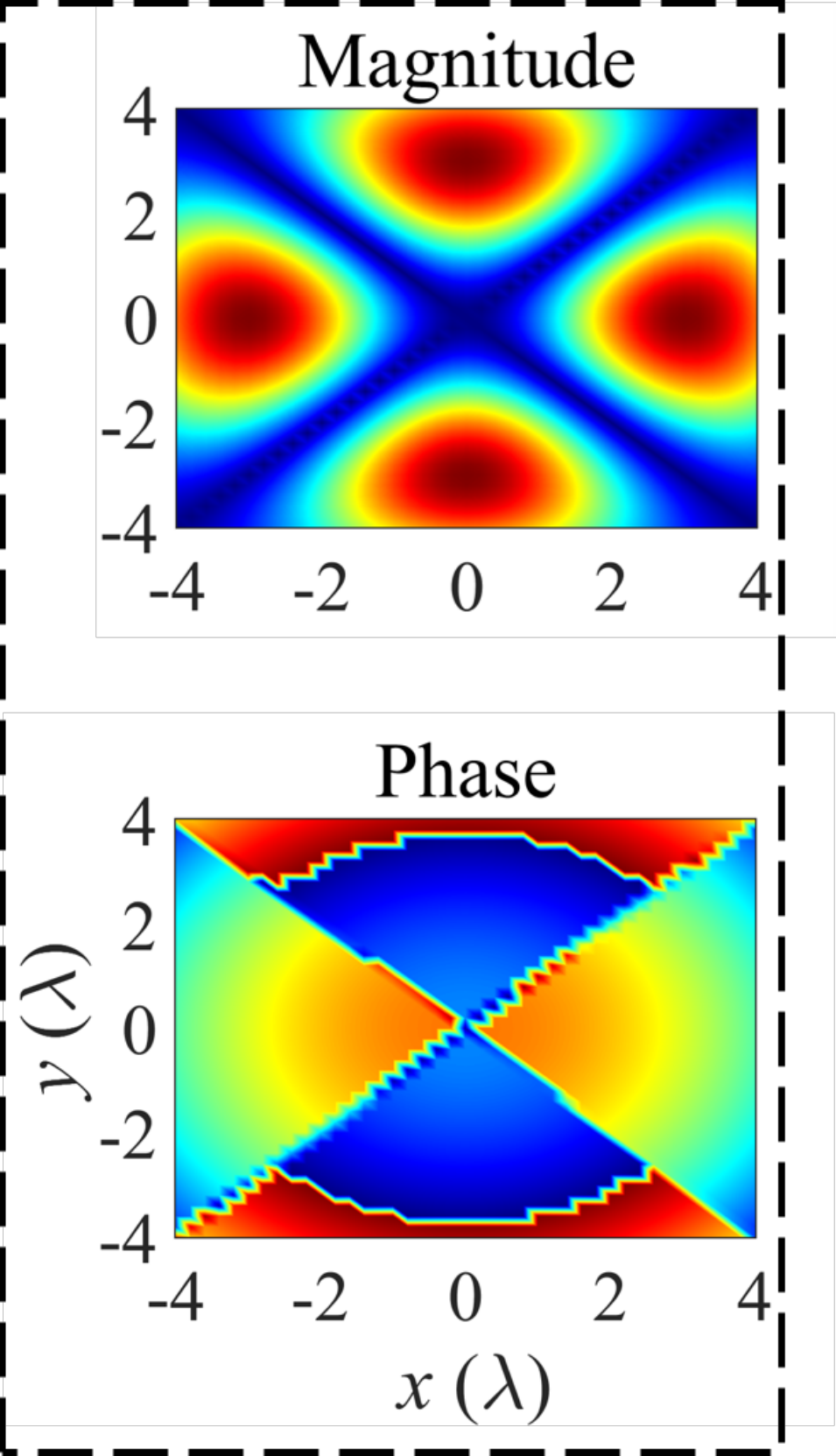}\label{fig7c}}%
\hfil
\begin{minipage}[b]{0.25\textwidth}
\centering
\subfloat[]{\includegraphics[width=1\textwidth,height=3.2cm]{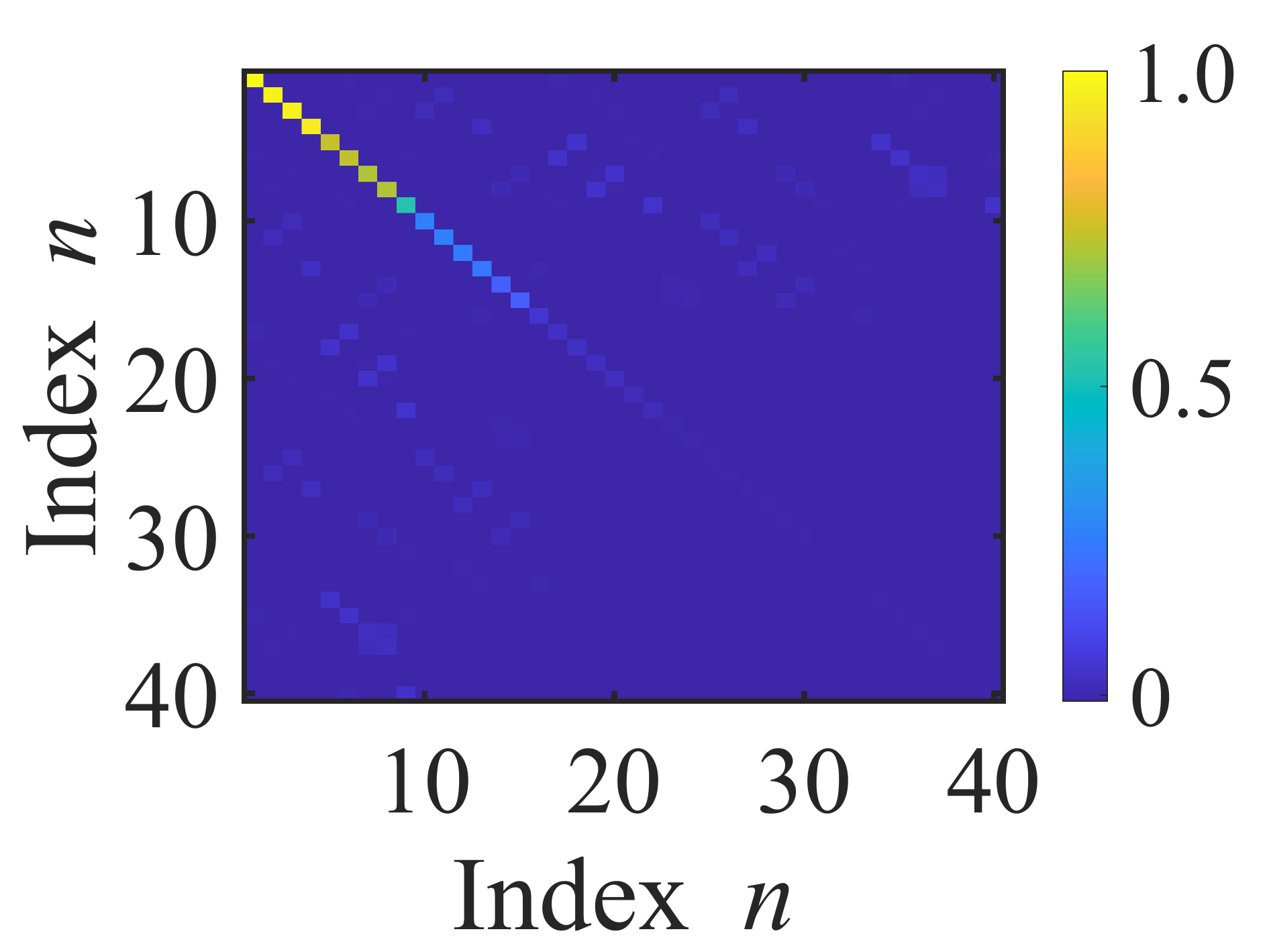}\label{fig7d}}
\vfill
\subfloat[]{\includegraphics[width=1\textwidth,height=3.2cm]{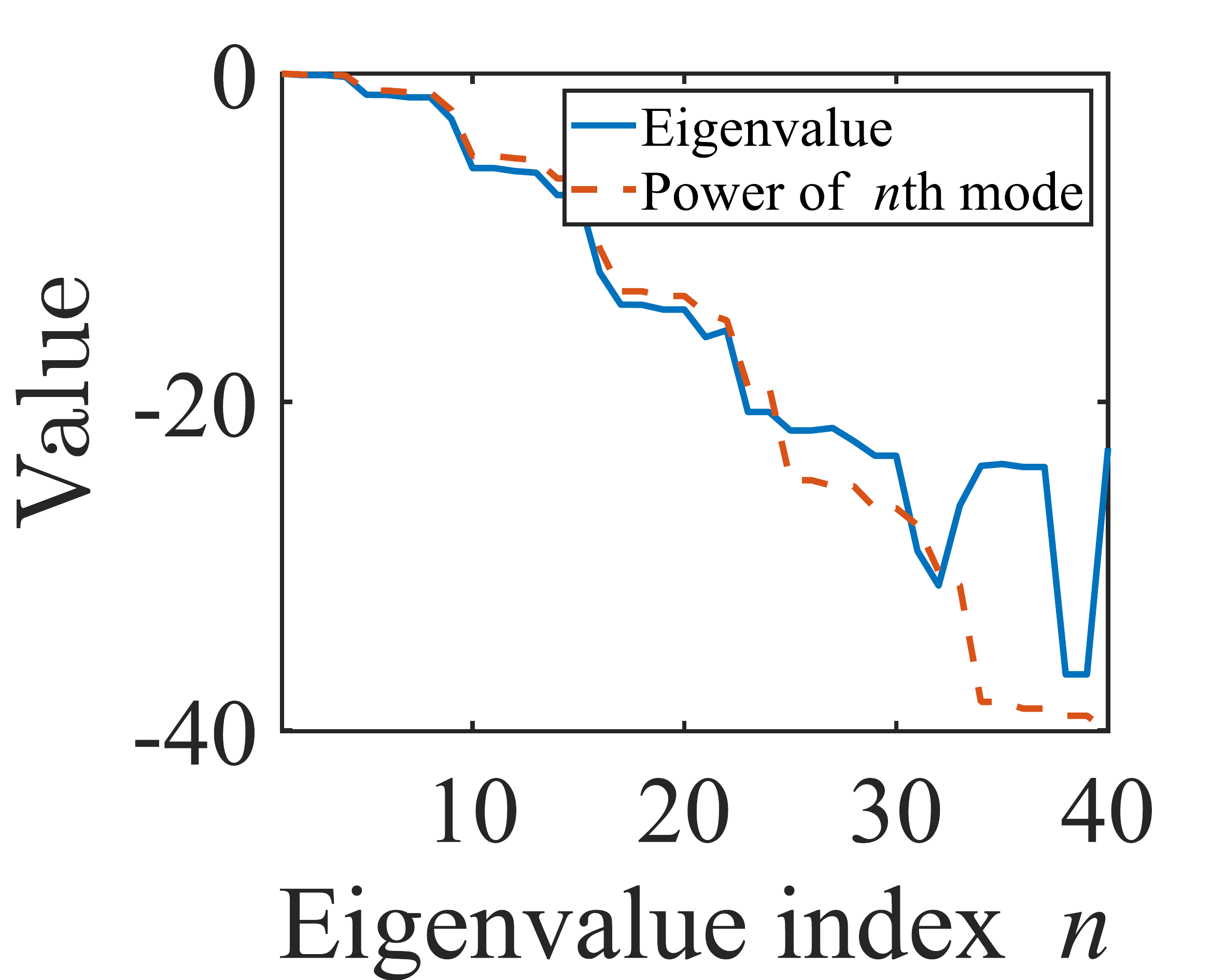}\label{fig7e}}
\end{minipage}
\caption{Illustration of the selected received fields radiated by the corresponding mode currents. (a) The magnitude and phase distributions of Field 1. (b) The magnitude and phase distributions of Field 3. (c) The magnitude and phase distributions of Field 5. (d) The correlation matrix of the first 40 received fields, which indicates the orthogonality of the fields. (e) The values on the diagonal of the correlation matrix for illustrating the normalized power of the nth field.}
\label{fig7}
\end{figure*}

We here present a series of the mode current (i.e., the eigenfunction) $\varphi_n(\mathbf{r})$ and the corresponding field $\psi_n(\mathbf{r})$ on the receiver, to validate the effectiveness of the numerical method and the orthogonality of these functions. As illustrative examples, Figs. \ref{fig6}\subref{fig6a}-\subref{fig6c} demonstrate the magnitude and phase distributions of mode currents $\varphi_1(\mathbf{r})$, $\varphi_3(\mathbf{r})$, and $\varphi_5(\mathbf{r})$ on the transmitter, respectively. Fig. \ref{fig6}\subref{fig6d} shows the correlation matrix of the first 40 mode currents, which is calculated by \eqref{eq12}. The results clearly show that the designed currents are orthonormal, which ensures the non-interference between different modes. \ref{fig6}\subref{fig6e} illustrates the values on the diagonal of the matrix, which quantifies the power of the nth mode current. Furthermore, Figs. \ref{fig7}\subref{fig7a}-\subref{fig7c} show the magnitude and phase distributions of the received fields launched by the mode currents, which are $\psi_1(\mathbf{r})$, $\psi_3(\mathbf{r})$, and $\psi_5(\mathbf{r})$, respectively. Fig. \ref{fig7}\subref{fig7d} presents the correlation matrix of the received fields, which is obtained by \eqref{eq14}. Furthermore, Fig. \ref{fig7}\subref{fig7e} demonstrates the values on the diagonal of the matrix wherein they quantify the power of the field on the receiver. It can be seen that, as shown in \eqref{eq14}, the power of the received field launched by the nth mode current is almost equal to the corresponding eigenvalue $\beta_n$. The small error here is mainly due to the numerical integration.

\begin{figure*}[!ht]
\centering
\subfloat[]{\includegraphics[width=0.52\textwidth]{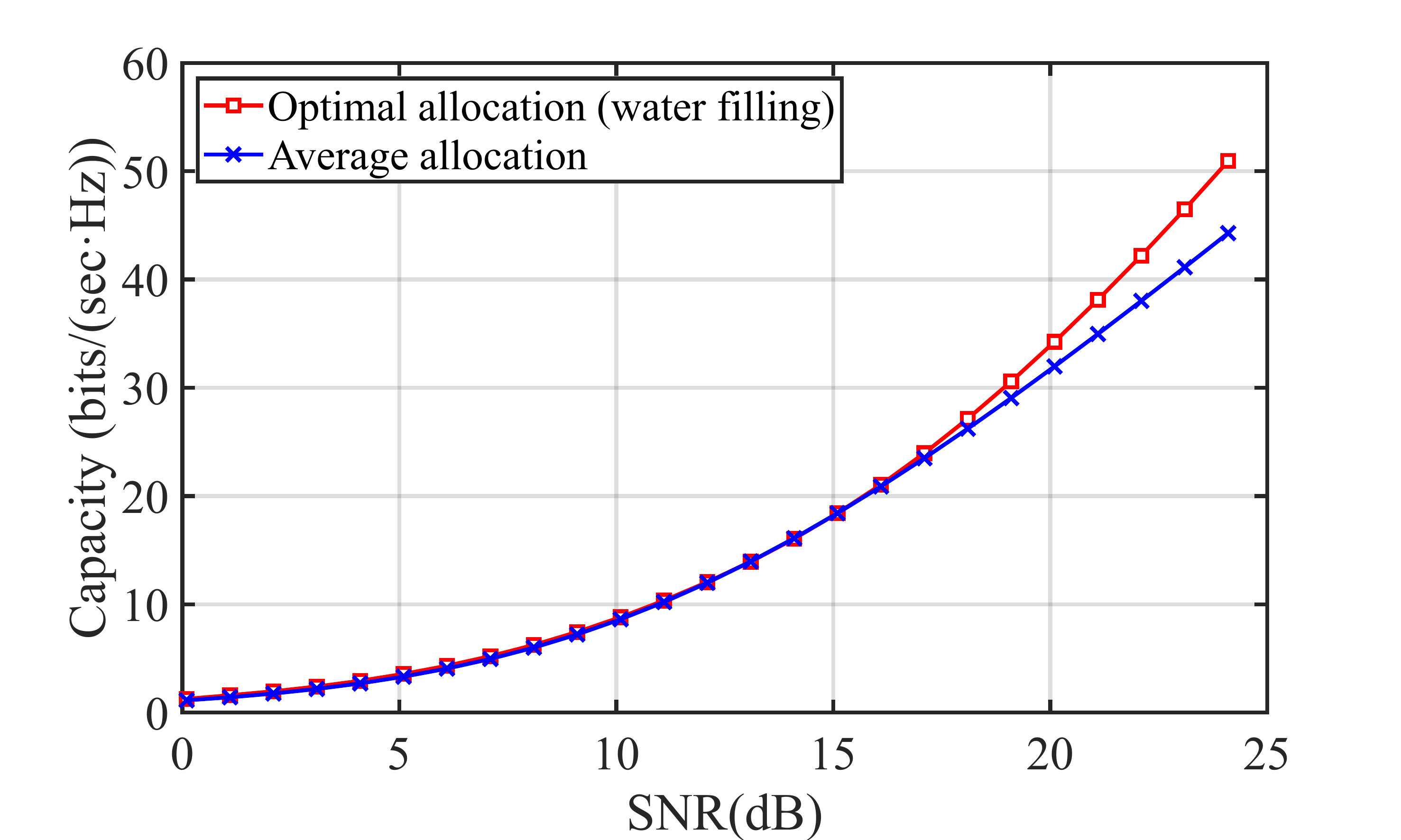}\label{fig8a}}
\hfill
\centering
\subfloat[]{\includegraphics[scale=0.34]{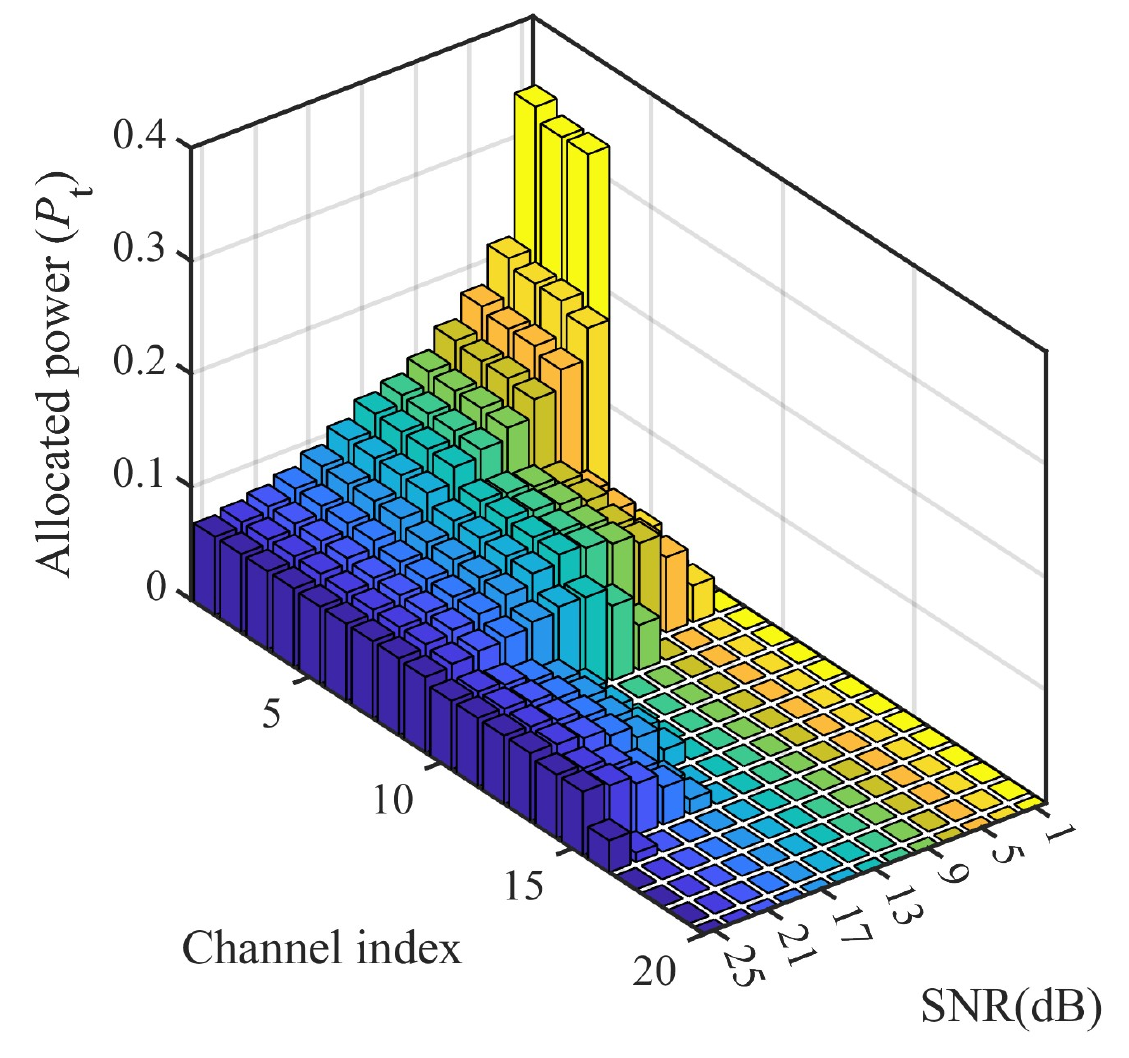}\label{fig8b}}
\caption{Illustration for understanding the relationship between the capacity proposed by the water-filling algorithm and the average allocated method. (a) The EM channel capacity as a function of SNR. (b) The allocated power of each subchannels as a function of SNR, obtained by the water-fillling algorithm.}
\label{fig8}
\end{figure*}

Once the complete channel information has been obtained, we undertake a detailed analysis of the system capacity bound, taking into full consideration the relationship between the spatial DoF, noise, and transmitted power. Fig. \ref{fig8}\subref{fig8a} illustrates the variation in EM channel capacity as a function of SNR, obtained through two distinct calculation methodologies. In the first methodology, the optimal capacity is obtained by water-filling algorithm, as shown in \eqref{eq25}. The other is based on the average power allocation of the first N subchannel, as formulated in \eqref{eq27}. It is observed that the capacity obtained by the average power allocation is nearly identical to that of the optimum in cases with the low (less than 5 dB) and the medium (5 to 15 dB) SNR regimes. For the high SNR regime, the capacity discrepancy between the two methods gradually increases. Fig. \ref{fig8}\subref{fig8b} demonstrates the power allocation of each subchannel under different SNRs, which helps to further explain the previous phenomenon. In the case of high SNR, the water-filling algorithm tends to allocate power equally among the sub-channels. In this case, the number of effective channels with non-zero allocated power exceeds the spatial DoF N given by \eqref{eq20}. It is worth noting that as N increases, the capacity calculated by \eqref{eq27} will increase gradually and eventually reach a bound. On the contrary, the number of effective channels for optimal allocation is less or equal to than the spatial DoF given by \eqref{eq20} in the low and medium SNR regimes, resulting in capacities obtained by \eqref{eq25} and \eqref{eq27} that are nearly identical.

\section{Conclusion}
In this paper, an EM channel capacity formula is deduced and analyzed, with the aim of exploring the spatial degree of freedom of EM waves. The formula provides an upper bound on the capacity of the transmitter and the receiver of given apertures and distance. Furthermore, a novel channel model that covers the vector EM waves and the scenes of near and far fields is constructed, and the characteristics of the finite angular spectral bandwidth is revealed. The optimal precoder and combiner are obtained by formulating the problem of the maximal power transmission between the transmitter and the receiver. The relationship between the spatial DoF, the noise, and the power is deeply analyzed. This work presents a novel perspective for understanding EIT and a new method for the design of pattern functions and the channel model for holographic communications. The future works will be focused on the phase quantization effect and expand the proposed theory to encompass the space-time-coding metasurfaces.

{\appendices
\section{Derivations that Transform Problem \eqref{eq8} to Problem \eqref{eq9}}\label{appA}
The explicit expression of the field power on the receiver is
\begin{equation}
\label{eq30}
    \begin{split}
        & P_r=\frac{1}{2 \eta} \int_{{\mathcal{A}_r}}|E(\mathbf{r})|^2 d \mathbf{r} \\
& =\frac{1}{2 \eta} \int_{{\mathcal{A}_r}} E(\mathbf{r}) \bar{E}(\mathbf{r}) d \mathbf{r} \\
& =\frac{1}{2 \eta} \int_{{\mathcal{A}_r}} \int_{{\mathcal{A}_s}}\left[-\frac{k \omega \mu}{16 \pi^2} \int_{S_{\theta_e}} e^{-jk\hat{\mathbf{k}}\cdot\hat{\mathbf{r}}_{qs}} \cdot \alpha\left(\hat{\mathbf{k}} \cdot \hat{\mathbf{r}}_{p q}\right) e^{-jk\hat{\mathbf{k}}\cdot\hat{\mathbf{r}}_{rp}} d \hat{\mathbf{k}}\right] J(\mathbf{s}) d \mathbf{s} \bar{E}(\mathbf{r}) d \mathbf{r} \\
& =\frac{1}{2 \eta} \int_{{\mathcal{A}_t}} \bar{J}(\mathbf{s}) \int_{{\mathcal{A}_t}} M\left(\mathbf{s},\,\mathbf{s}^{\prime}\right) J\left(\mathbf{s}^{\prime}\right) d \mathbf{s}^{\prime} d \mathbf{s} \\
& \leq \frac{1}{2 \eta} \sqrt{\int_{{\mathcal{A}_t}}|\bar{J}(\mathbf{s})|^2 d \mathbf{s}} \cdot \sqrt{\int_{{\mathcal{A}_t}}\left|\int_{{\mathcal{A}_t}} M\left(\mathbf{s},\,\mathbf{s}^{\prime}\right) J\left(\mathbf{s}^{\prime}\right) d \mathbf{s}^{\prime}\right|^2 d \mathbf{s}} \\
& =\frac{\beta}{2 \eta} \int_{{\mathcal{A}_t}}|J(\mathbf{s})|^2 d \mathbf{s},
    \end{split}
\end{equation} where $M(\mathbf{s},\,\mathbf{s}^\prime)$ is the integral kernel, which is given by
\begin{equation}
\label{eq31}
M(\mathbf{s},\,\mathbf{s}')=\int_{{\mathcal{A}_r}}H(\mathbf{r},\,\mathbf{s}){\overline{H}}(\mathbf{r},\,\mathbf{s}')d\mathbf{r},
\end{equation} where $H(\mathbf{r,\,s})$ is
\begin{equation}
\label{eq32}
    H(\mathbf{r},\,\mathbf{s})=-\frac{k\omega\mu}{16\pi^2}\int_{S_{\theta_e}}e^{-jk\hat{\mathbf{k}}\cdot\hat{\mathbf{r}}_{qs}}\cdot\alpha(\hat{\mathbf{k}}\cdot\hat{\mathbf{r}}_{pq})e^{-jk\hat{\mathbf{k}}\cdot\hat{\mathbf{r}}_{rp}}d\hat{\mathbf{k}}.
\end{equation}

In \eqref{eq30}, we apply Cauchy-schwarz inequality, namely, $\int f(x)g(x)dx\leq\sqrt{\int| f(x)|^2dx}\cdot\\\sqrt{\int\mid g(x)\mid^2dx}$, and the equality holds when $g(x)=\beta\overline{f}(x)$. Therefore, the design of the current function supporting maximal power transmission can be equivalently reformulated as
\begin{equation}
\label{eq33}
\beta\varphi(\mathbf{s})=\int_{{\mathcal{A}_t}}M(\mathbf{s},\,\mathbf{s}')\varphi(\mathbf{s}')d\mathbf{s}',
\end{equation} wherein we replace the current character $J(\mathbf{s})$ with $\varphi(\mathbf{s})$, since the solution of \eqref{eq33} will generate a series of eigenvalue and eigen function pairs $(\beta_n,\,\varphi_n(\mathbf{s}))$. From \eqref{eq30}, we can see that the eigenvalue $\beta_n$ quantifies the power level of the fields launched by the mode current $\varphi_n(\mathbf{s})$. Finally, to meet the constraint of the transmitted power, the eigen function is normalized as
\begin{equation}
\label{eq34}
\varphi_n(\mathbf{s}):=\sqrt{P_t/\eta}\varphi_n(\mathbf{s}).
\end{equation}
\section{The Proof of the Properties of the Eigenvalues and Eigenfunctions}\label{appB}
\subsection{The Proof of Property (1)}
\begin{equation}
\label{eq35}
    \begin{split}
\beta_n\int_{{\mathcal{A}_t}}|\varphi_n(\mathbf{s})|^2 d\mathbf{s}& =\int_{{\mathcal{A}_t}}\bar{\varphi}_n(\mathbf{s})\int_{{\mathcal{A}_t}}M(\mathbf{s},\,\mathbf{s}')\varphi_n(\mathbf{s}')d\mathbf{s}'d\mathbf{s} \\
&\overset{(a)}{=}\int_{{\mathcal{A}_t}}\varphi_n(\mathbf{s}^{\prime})\int_{{\mathcal{A}_t}}\bar{M}(\mathbf{s}^{\prime},\,\mathbf{s})\overline{\varphi}_n(\mathbf{s})d\mathbf{s}d\mathbf{s}^{\prime} \\
&=\overline{\beta}_n\int_{{\mathcal{A}_t}}|\varphi_n(\mathbf{s}')|^2 d\mathbf{s}'.
\end{split}
\end{equation} In (35a), we switch the integral order and utilize the symmetric feature of the kernel $M(\mathbf{s},\,\mathbf{s}')$ . Considering the non-zero solution of the eigenfunctions, we obtain $\beta_n=\hat{\beta}_n$, showing that the eigenvalue $\beta_n$ is real.
\subsection{The Proof of Property (2)}
\begin{equation}
\label{eq36}
    \begin{split}
\int_{{\mathcal{A}_t}}\varphi_m(\mathbf{s})\overline{\varphi}_n(\mathbf{s})d\mathbf{s}& =\int_{{\mathcal{A}_t}}\frac{1}{\beta_m}\int_{{\mathcal{A}_t}}M(\mathbf{s}, \mathbf{s}')\varphi_m(\mathbf{s}')d\mathbf{s}'\overline{\varphi}_n(\mathbf{s})d\mathbf{s} \\
&\overset{(a)}{=}\frac1{\beta_m}\int_{{\mathcal{A}_t}}\varphi_m(\mathbf{s}^{\prime})\int_{{\mathcal{A}_t}}\bar{M}(\mathbf{s},\mathbf{s}^{\prime})\overline{\beta}_n(\mathbf{s})d\mathbf{s}d\mathbf{s}^{\prime} \\
&=\frac{\beta_n}{\beta_m}\int_{{\mathcal{A}_t}}\varphi_m(s')\overline{\varphi}_n(s')ds',
\end{split}
\end{equation} where (36a) employs the switching of the order of integration and the symmetric feature of the kernel $H(\mathbf{s,\,s}')$. Then the above equation is $(\beta_m-\beta_n)/\beta_m\int_{\mathcal{A}_t}\varphi_m(\mathbf{s})\overline{\varphi}_n(\mathbf{s})d\mathbf{s}=0.$ Hence, $\int_{\mathcal{A}_t}\varphi_m(s)\overline{\varphi}_n(s)ds=\delta_{mn}.$

\subsection{The Proof of Property (3)}
Using the Parseval’s theorem, we have
\begin{equation}
\label{eq37}
\int_{\mathcal{A}_t}\varphi_m(\mathbf{s})\overline{\varphi}_n(\mathbf{s})d\mathbf{s}=\int_{\mathcal{A}_{sum}}\psi_m(\mathbf{r})\overline{\psi}_n(\mathbf{r})d\mathbf{r},
\end{equation} therefore, $\int_{\mathcal{A}_{\mathrm{sum}}}\psi_m(\mathbf{r})\bar{\psi}_n(\mathbf{r})d\mathbf{r}=\delta_{mn}$, showing that $\psi_n(\mathbf{r})$ is orthonormal on $\mathcal{A}_{sum}$.

\begin{equation}
\label{eq38}
    \begin{aligned}
\int_{\mathcal{A}_r}\psi_m(\mathbf{r})\bar{\psi}_n(\mathbf{r})d\mathbf{r}& =\int_{\mathcal{A}_r}\int_{\mathcal{A}_t}H(\mathbf{r,s})\varphi_m(\mathbf{s})d\mathbf{s}\int_{\mathcal{A}_t}\bar{H}(\mathbf{r,s})\overline{\varphi}_m(\mathbf{s})d\mathbf{s}d\mathbf{r} \\
&=\int_{\mathcal{A}_t}\overline{\varphi}_n(\mathbf{s})\int_{\mathcal{A}_t}M(\mathbf{s}, \mathbf{s}')\varphi_m(\mathbf{s}')d\mathbf{s}'d\mathbf{s} \\
&=\beta_m\int_{\mathcal{A}_t}\varphi_m(\mathbf{s})\bar{\varphi}_n(\mathbf{s})ds \\
&=\beta_m\delta_{mn},
\end{aligned}
\end{equation} indicating that $\psi_m(\mathbf{r})$ is orthogonal on the receiver.

\section{Derivations of \eqref{eq18}} \label{appC}
Substituting \eqref{eq15} into \eqref{eq1}, and considering \eqref{eq13} and \eqref{eq16}, we then have
\begin{equation}
    \begin{split}
E(\mathbf{r})&=-j\omega\mu\int_{\mathcal{A}_{t}}G(\mathbf{r}, \mathbf{s})J(\mathbf{s})d\mathbf{s} \\
&\overset{(a)}{\operatorname*{=}}\sum_{n=1}^\infty x_n\int_{\mathcal{A}_t}H(\mathbf{r}, \mathbf{s})\varphi(\mathbf{s})d\mathbf{s} \\
&=\sum_{n=1}^\infty x_n\psi_n(\mathbf{r}) \\
&=\sum_{n=1}^\infty\sqrt{\beta_n}x_n\chi_n(\mathbf{r}).
\end{split}
\end{equation} In (39a), we replace $-j\omega\mu \mathbf{\bar{G}(r,\,s)}$ with $H(\mathbf{r,\,s})$. Considering the representation of the noise (i.e., \eqref{eq17}), the received field is expressed as
\begin{equation}
\label{eq40}
E_n(\mathbf{r})=\sum_{n=1}^\infty[\sqrt{\beta_n}x_n+\sigma_n]\chi_n(\mathbf{r}).
\end{equation}

\section{The Numerical Method for Solving Problem \eqref{eq9}} \label{appD}
To analyze arbitrary source and observable regions, it is necessary to develop numerical methods that provide approximate solutions to the eigenfunctions. In this section, we represent the eigenfunction by a linear combination of sub-basic functions, namely,
\begin{equation}
\label{eq41}
    \varphi_n(\mathbf{s})=\sum_ia_{ni}e_i(\mathbf{s}),
\end{equation}
where $\{a_{ni}\}_{i=1}^{\infty}$ are the expansion coefficients, and $\{e_{i}\}_{i=1}^{\infty}$ are the known and complete orthonormal bases on the transmitter. Then the integral equation can be expressed as
\begin{equation}
\label{eq42}
    \begin{split}
\beta_n\sum_ia_{ni}e_i(\mathbf{s})& =\int_{\mathcal{A}_t}M(\mathbf{s}, \mathbf{s}')\sum_ia_{ni}e_i(\mathbf{s}')d\mathbf{s}' \\
&=\sum_ia_{ni}(\sum_jb_{ji}e_j(\mathbf{s})) \\
&\overset{(a)}{=}\sum_i(\sum_ja_{nj}b_{ij})e_i(\mathbf{s}),
\end{split}
\end{equation} where

\begin{equation}
\label{eq43}
    b_{ij}=\int_{\mathcal{A}_t}\overline{e}_i(\mathbf{s})\int_{\mathcal{A}_t}M(\mathbf{s},\,\mathbf{s}^{\prime})e_j(\mathbf{s}^{\prime})d\mathbf{s}^{\prime}d\mathbf{s},
\end{equation}
and in (42a), we switch the sum order. Since the $e_i(\mathbf{s})$ is orthonormal, we then obtain
\begin{equation}
\label{eq44}
    \beta_na_{ni}=\sum_ja_{nj}b_{ij},\quad i=1, 2, ...
\end{equation} Considering all expansion coefficients, equation \eqref{eq44} can be transformed to the matrix form, namely,
\begin{equation}
\label{eq45}
    \beta\mathbf{a}=\mathbf{Ba},
\end{equation} where $\beta$ and $\mathbf{a}$ are the eigenvalue and eigenvector of the matrix $\mathbf{B}$, respectively. Their explicit expressions are
\begin{equation}
\label{eq46}
    \mathbf{a}=(a_1,\,a_2,\,...)^T,\quad\mathbf{B}=\begin{pmatrix}b_{11}&b_{12}&\cdots\\b_{21}&b_{22}&\cdots\\\cdots&\cdots&\cdots\end{pmatrix}.
\end{equation} Therefore, the solution of the problem \eqref{eq9} is transformed into the eigen decomposition of the matrix $\mathbf{B}$. It is worth noting that the element $b_{ij}$ can perform offline, which effectively reduces the computational complexity. Furthermore, we artificially truncate the matrix $\mathbf{B}$ to obtain the approximate numerical solution. More details about the method of construction and truncation of the matrix $\mathbf{B}$ are shown in Appendix E.

\section{The Details on the Realization of the Numerical Method} \label{appE}
As a case study, we use a rectangular aperture array $\boldsymbol{\mathcal{A}_t}:L_x\times L_y$ to illustrate the details of the numerical method outlined in Appendix D. We firstly show how to construct the sub-basic functions $e_i(\mathbf{s})$.The standard Legendre function constitutes a complete orthogonal set on the interval $[-1,\,1]$ and exhibits many good properties, which is easily applied to a 2-D square region through a straightforward extension. According to the following three-term recursion formula
\begin{equation}
\label{eq47}
    P_n(x)=\frac{2n-1}nxP_{n-1}(x)-\frac{n-1}nP_{n-2}(x),\quad n\geq2,
\end{equation} and the first two terms
\begin{equation}
\label{eq48}
    P_0(x)=1,\quad P_1(x)=x,
\end{equation}
we can get the 1-D Legendre function of any order. The orthogonal relation between Legendre functions of different orders is
\begin{equation}
\label{eq49}
    \int_{-1}^1P_m(x)P_n(x)dx=\frac2{2n+1}\delta_{mn}.
\end{equation}
To construct the orthonormal basis on the interval $[-L_x/2,\,L_x/2]$, we perform the affine transformation on Legendre functions, namely,
\begin{equation}
\label{eq50}
    e_n(x)=\sqrt{\frac{2n+1}{L_x}}P_n(\frac{2x}{L_x}),\quad-L_x/2\leq x\leq L_x/2.
\end{equation}
We then obtain the 2-D orthonormal basis on the square region $\boldsymbol{\mathcal{A}_t}:L_x\times L_y$, specifically,
\begin{equation}
\label{eq51}
    \begin{aligned}e_{mm}(\mathbf{s})&=e_m(x)e_n(y)\\&=\sqrt{\frac{(2m+1)(2n+1)}{L_xL_y}}P_m(\frac{2x}{L_x})P_n(\frac{2y}{L_y}), \\
    &-L_x/2\leq x\leq L_x/2,\, -L_y/2\leq y\leq L_y/2,
    \end{aligned}
\end{equation} where $\mathbf{s}=\left(x,\,y\right)^T$.

The next step is to determine the total number of the orthonormal basis, which is also the truncation of the matrix $\mathbf{B}$. We can find that the expression of the 2-D basis depends on two variables, i.e., the orders along the $x$ and $y$ directions. To decrease the variable, we construct the $i$th sub-basic function by the following rules, namely,

\begin{equation}
\label{eq52}
    \begin{aligned}&e_i(\mathbf{s})=e_i^j(\mathbf{s})=e_m(x)e_n(y),\\
    &0\leq m, n\leq j;\,m+n=j; j=0,\,1,\,...,\,t,\end{aligned}
\end{equation} where $m$ and $n$ are the orders of the 1-D orthonormal function along the $x$ and $y$ directions, respectively. $j$ is the total order along the $x$ and $y$ directions. $t$ is the maximal order of the 2-D orthonormal function. According to the rule defined by \eqref{eq52}, the sub-basic function only depends on the maximal order $t$ and the total number of basic functions is $N=(t+1)(t+2)/2$. 

Table \ref{tab1} presents a representative example to illustrate the construction of the 2-D orthonormal function. Specifically, the maximal order is $t=3$, and thus the total number of basic functions is $N=10$. The pseudo-code of the numerical method for arbitrary source and observable regions is summarized in Algorithm \ref{alg1}. 
\begin{table}[!ht]
\renewcommand{\arraystretch}{1.25} 
\setlength{\tabcolsep}{25pt} 
\captionsetup{font=footnotesize,justification=centering}
\caption{The Expression of the First 10 Sub-basic Functions, as a Representative Example to Illustrate the Construction Rule.}
\label{tab1}
    \centering
    \scalebox{1.0}{\begin{tabular}{c c c c}
        \toprule
          & $i$ & Expression & Total order: $j$ \\
          \midrule
          $e_1^0(\mathbf{s})$ & 1 & $e_0(x)e_0(y)$ & 0 \\
          $e_2^1(\mathbf{s})$ & 2 & $e_0(x)e_1(y)$ & \multirow{2}{*}{1} \\
          $e_3^1(\mathbf{s})$ & 3 & $e_1(x)e_0(y)$ \\
          $e_4^2(\mathbf{s})$ & 4 & $e_0(x)e_2(y)$ & \multirow{3}{*}{2} \\
          $e_5^2(\mathbf{s})$ & 5 & $e_1(x)e_1(y)$ \\
          $e_6^2(\mathbf{s})$ & 6 & $e_2(x)e_0(y)$ \\
          $e_7^3(\mathbf{s})$ & 7 & $e_0(x)e_3(y)$ & \multirow{4}{*}{3} \\
          $e_8^3(\mathbf{s})$ & 8 & $e_1(x)e_2(y)$ \\
          $e_9^3(\mathbf{s})$ & 9 & $e_2(x)e_1(y)$ \\
          $e_{10}^3(\mathbf{s})$ & 10 & $e_3(x)e_0(y)$ \\
          \bottomrule     
    \end{tabular}}
\end{table}

\begin{algorithm}[ht]
    \caption{The Numerical Solution of Problem \eqref{eq9}}
    \label{alg1}
    \renewcommand{\algorithmicrequire}{\textbf{Input:}}
    \renewcommand{\algorithmicensure}{\textbf{Output:}}
    \begin{algorithmic}[1]
        \REQUIRE The transmitter $\boldsymbol{\mathbf{A}_t}$, the receiver $\boldsymbol{\mathbf{A}_r}$, and the maximal order $t$  
        \ENSURE The eigenvalue $\beta_n$ and the eigenfunction $\varphi_n(\mathbf{s})$. 
        
        \STATE  Choose the approximate sub-basic function according to the shape of the source $\boldsymbol{\mathbf{A}_t}$.
        \STATE  Orthonormalize the sub-basic function using \eqref{eq50}.
        \FOR{$i=1$ To $(t+1)(t+2)/2$}
            \STATE  Construct the $i$th 2-D orthonormal function $e_i(\mathbf{s})$ using \eqref{eq51}.
        \ENDFOR
        \FOR{$i=1$ To $(t+1)(t+2)/2$}
            \FOR{$j=1$ To $(t+1)(t+2)/2$}
                \STATE  Construct $b_{ij}$ using \eqref{eq43}.
            \ENDFOR
        \ENDFOR
        \STATE  Eigen decomposition the matrix B to obtain the eigenvalue and eigenvector pairs $(\beta_n,\,\mathbf{a}_n)$.
        \STATE  Construct the $n$th eigenfunction using $\varphi_n(\mathbf{s})=\sum\limits_{i=1}^{(t+1)(t+2)/2}a_{ni}e_i(\mathbf{s})$.
    \end{algorithmic}
\end{algorithm}

\vfill

\end{document}